\documentclass[11pt,x11names,a4paper]{article}

\usepackage{multirow}
\usepackage[utf8]{inputenc}
\usepackage{lmodern}
\usepackage[T1]{fontenc} 
\usepackage{microtype} 

\usepackage[a4paper, left=25mm, right=25mm, top=30mm, bottom=25mm]{geometry} 

\usepackage{abstract}

\usepackage{xcolor}

\usepackage{cite}
\usepackage{hyperref}
\hypersetup{
	colorlinks=true,
	linkcolor=Blue4,
	citecolor=Red4,
	urlcolor=Green4,
	linktoc=page
}

\usepackage{tikz}
\usetikzlibrary{decorations.markings,positioning}

\usepackage{amsmath,amssymb,slashed,mathbbol,mathtools}
\numberwithin{equation}{section}

\title{\fontsize{20pt}{24pt}\selectfont\textbf{Holographic Ensembles in Type IIB}\vspace{2mm}}

\author{
\large{ \href{mailto:jvanmuid@ic.ac.uk}{Jesse van Muiden}}\\[5mm]
{}{Abdus Salam Centre for Theoretical Physics, Imperial College London}\\
{\normalsize Prince Consort Road, London SW7 2AZ, UK}\\[5mm]
}

\date{}

\usepackage{booktabs}

\newcommand{\rmi}{\mathrm{i}}
\newcommand{\rme}{\mathrm{e}}
\newcommand{\rmd}{\mathrm{d}}

\newcommand{\Z}{\mathbf{Z}}
\newcommand{\UU}{\mathrm{U}}

\newcommand{\Xip}{\widehat\Xi_{\rm pert}}
\newcommand{\Xie}{\widehat\Xi_{\rm exact}}

\newcommand{\Jp}{J_{\rm pert}}
\newcommand{\Je}{J_{\rm exact}}
\newcommand{\mueff}{\mu_{\rm eff}}

\newcommand{\Aa}{\mathcal{A}}
\newcommand{\Akk}{\mathcal{A}_{\rm KK}}

\begin{document}
{\hypersetup{urlcolor=black}\maketitle}
\thispagestyle{empty}

\begin{abstract}
\noindent We study holographic ensembles in type IIB string theory on the Schur twisted AdS$_5 \times S^5$ background, and its corresponding boundary partition function: the Schur index of $\mathcal N=4$ SYM. In the supergravity limit the two natural ensembles are either at fixed flux on $S^5$, corresponding to a boundary theory at fixed rank, or its conjugate choice of fixed potential on the boundary of AdS$_5$, corresponding on the boundary to a grand canonical ensemble of theories at fixed chemical potential. Due to the self-dual nature of the five-form in type IIB supergravity the choice of ensembles is not seen in the standard bulk action, but rather by the coupling of an additional topological term. We compute the bulk partition function to one loop, including the Kaluza-Klein determinant and the measure of the ensemble-changing sum, and show that depending on the sign of this topological term it matches the perturbative boundary partition function in either the canonical or the grand canonical ensemble.
\end{abstract}

\section{Introduction}
Conventionally we study top-down holography \cite{Maldacena:1997re,Gubser:1998bc,Witten:1998qj} at fixed rank of the boundary gauge group, $N$. Recently, it has been proposed that it might be easier or even more natural in the bulk to instead let $N$ fluctuate and fix its conjugate chemical potential $\mu$ \cite{Gautason:2025plx}. The proposal arose from precision studies in holography with the aim to fully match the bulk and boundary partition functions in M-theory
\begin{equation}
	\mathcal Z_{\text{M2}}[g,A_3,\Psi] = \log Z_{\text{QFT}}[J]
\end{equation}
where the left hand side is to be decomposed into a perturbative supergravity part and non-perturbative membrane instantons which are quantised at fixed background source $A_3$ \cite{Gautason:2025per}.\footnote{The study of semi-classically quantising these holographic membrane instantons was first pursued in \cite{Beccaria:2023ujc} following \cite{Gautason:2023igo}.} To find a match one is to correctly identify the boundary conditions and background sources. Since the bulk is computed at fixed source $A_3$ one is to fix boundary conditions for this background field and identify the corresponding dual source. In the context of 3d holographic CFTs dual to M-theory on AdS$_4 \times M_7$ this source is exactly the chemical potential in the grand canonical ensemble of the boundary theory. To compute observables in the canonical ensemble of the boundary theory, at fixed rank $N$, one has to transform to an ensemble of fixed $\star G_4$ afterwards. The details of this transform were the focus of interest in \cite{Bobev:2026gir}, where it was shown that the measure and contour of the integral are determined by the topology of the asymptotic boundary in AdS$_4$.

To verify this approach it is important to have a good control of the holographic setup and perform non-trivial precision tests, both in the canonical and grand canonical ensemble of the boundary theory. In AdS$_4$ the prime example is the ABJM theory \cite{Aharony:2008ug} and its sphere partition function which was in fact studied most easily through its grand canonical ensemble, e.g. in \cite{Marino:2011eh,Hatsuda:2012hm} and follow up works. One of the advantages of this theory is that its (supersymmetrically localised) partition function was mapped to a theory of free fermions, which was studied at both fixed particle number and fixed chemical potential. In the bulk description it was shown in \cite{Gautason:2025plx} that the leading two-derivative (eleven-dimensional) supergravity action corresponded to the leading term of the grand potential at large chemical potential in the boundary theory, and importantly does not match the large-$N$ answer of the canonical partition function. This has long been overlooked due to a four-dimensional gauged supergravity approach in which the gravitational coupling was fixed to match the boundary answer (which on the sly fixes a particular set of boundary terms). A particularly interesting consequence, which can be empirically deduced from the boundary grand potential, is that in these supersymmetric backgrounds the bulk partition function truncates at the eighth derivative order supergravity action and instead the well-known Airy function \cite{Marino:2011eh,Fuji:2011km}, as the boundary partition function, comes from the ensemble changing integral\footnote{We have left out the $\mu^0$ contribution which is not coming from a local bulk contribution but instead from a one-loop term. From a worldvolume point of view it is to be expected to come from the constant map contribution, a degenerate membrane with momentum along the M-theory circle.}
\begin{equation}\label{Eq: Airy integral}
	\mathcal Z_{\text{M2}}(\mu) = \frac{\mathcal C}{3} \mu^3 + \mathcal B \mu \quad \Rightarrow \quad \frac{1}{\mathcal C^{1/3}}\text{Ai}\left[\frac{N - \mathcal B}{\mathcal C^{1/3}}\right] \sim \frac{1}{2\pi\rmi}\int_{-\pi \rmi}^{\pi\rmi} \rmd \mu \, \sum_{n\in \mathbf Z} \rme^{\mathcal Z_{\text{M2}}(\mu+2\pi\rmi n) - \mu N}\,.
\end{equation}
For boundaries with non-trivial topology the ensemble changing integral, which effectively integrates out $A_3$, is more involved and its measure can in that case be determined through a counting of zero-modes \cite{Bobev:2026gir}. For black hole geometries the fixed flux partition function is more involved and becomes a product of Airy functions or even sums of products of Airy functions, due to these zero-modes whose counting is dependent on the bulk topology and changes the measure of the integral.\footnote{In \cite{BenettiGenolini:2026cyc} it was proposed that for spherical black holes a product of Airy functions in the fixed flux ensemble is to arise as a product of integral representations of Airy function, as in \eqref{Eq: Airy integral}, which correspond to integrals over different equivariant parameters in the bulk.} This counting of zero-modes is in fact ensemble dependent and even for the sphere partition function there is a difference as the bulk partition function does not have a log contribution, contrary to the analysis in \cite{Bhattacharyya:2012ye}.

As can be seen from the discussion so far the study of ensembles in top-down holography has been tested rigorously in the context of M-theory and asymptotically AdS$_4$ geometries. And it has shown that the simple procedure of changing ensembles, or boundary conditions, resums infinitely many $1/N$ corrections into a particularly simple cubic polynomial. This resummation itself is thus a consequence of changing boundary conditions, not of supersymmetry, although it is to be expected that the simple looking form of the partition function in the grand canonical ensemble is a result of that supersymmetry, and it would be very interesting to derive this rigorously. 

It is an interesting question if we can use such resummations to our advantage also in string theory. Upon a reduction from M-theory to the type IIA string we can directly apply the same ensemble resummation which in this case corresponds to resumming infinitely many $g_s$ and $\alpha'$ corrections. In this paper we focus instead on the IIB string which does not have a geometrization of the string coupling. The canonical setup to start with is the IIB string on AdS$_5 \times S^5$ and its boundary $\mathcal N=4$ SYM theory, equating
\begin{equation}
	\mathcal Z_{\text{IIB}}[g,B_2,\ldots] = \log Z_{\mathcal N=4}[J],
\end{equation}
where the background sources need to be identified correctly once again, and the string partition function expands at low energies into the IIB supergravity action plus corrections
\begin{equation}
	\mathcal Z_{\text{IIB}}[g,B_2,\ldots] = -S_{\text{IIB}}[g,B_2,\ldots] + \ldots 
\end{equation}
To be able to test the holographic correspondence in different ensembles we again need a well-controlled case study in which both the canonical and grand canonical ensembles of the field theory are under good control. In the context of the $\mathcal N=4$ theory with gauge group $\text{U}(N)$ one of the simplest examples is the Schur index \cite{Gadde:2011ik,Gadde:2011uv}, which was shown to have an ideal Fermi-gas description, just as the sphere partition function of the ABJM theory, in \cite{Bourdier:2015wda}. And once again, it turned out more fruitful to study the grand canonical partition function of the theory, instead of the canonical one. Our aim is to understand the bulk description of this grand canonical ensemble and make progress in the context of holographic ensembles just as in the case of AdS$_4$ geometries in M-theory. This type IIB setup is however more subtle than the M-theory one described above, with the main reason being the self-dual constraint on the five-form in IIB supergravity. Canonically one equates and fixes the integrated five-form flux on the internal $S^5$ to the dual rank $N$ and works in the canonical ensemble on the boundary. To work instead in the grand canonical ensemble one fixes the electric component $C_4^{\mathrm{el}}$ at the AdS$_5$ boundary with Dirichlet conditions. Since the flux is imposed to be self-dual by hand this is a subtle procedure. The resolution is to work in the PST formalism, splitting the electric and magnetic parts of the five form flux, and supplement the action with a topological term \cite{Kurlyand:2022vzv}. The addition of this topological term does not alter the supergravity equations of motion, and thus is invisible to the worldsheet $\beta$-functions, but does alter the action and its on-shell action and ultimately needs to be derived directly from the quantisation of a degenerate genus zero string as in \cite{Fradkin:1985fq,Fradkin:1984pq,Tseytlin:1988tv}. An additional subtlety to keep in mind is that the boundary theory was studied with gauge group $\text{U}(N)$, in the string partition function we must therefore include the boundary degrees of freedom of the doubleton \cite{Gunaydin:1984fk,Kim:1985ez,Aharony:1998qu}, which we do throughout. 

We find that the coupling of this additional topological term can be chosen in two consistent ways which in fact determines to which boundary ensemble the bulk corresponds, similar to the ensembles that were found in M-theory. It would be very interesting to understand these ensembles directly from a worldsheet computation which fixes the couplings in the effective supergravity theory, including the topological contributions. Computing the bulk partition function in the twisted AdS$_5 \times S^5$ background to one loop, including the on-shell action, the Kaluza–Klein determinant, and the measure of the ensemble-changing sum we find a match for the perturbative contributions in both the canonical and grand canonical ensembles, with the correct bulk couplings for the topological term. We review the remaining non-perturbative contributions to the grand canonical partition function and show that they resum in an interestingly simpler form than in the canonical ensemble where they have been argued to arise from D3-brane giant gravitons \cite{Gaiotto:2021xce,Imamura:2021ytr,Arai:2020qaj,Beccaria:2024vfx,Gautason:2024nru}.

Contrary to M-theory, where there is a natural choice of ensemble since the supergravity action is written naturally in terms of the three-form gauge potential, in the IIB theory there is no preferred ensemble and instead the boundary polarisations are determined by the sign of the additional topological term in the IIB supergravity action, which remains to be derived from a worldsheet calculation. 
\section{Ensembles in IIB}
As mentioned in the introduction our case study for the analysis of ensembles in IIB string theory is the $Y = \text{AdS}_5 \times S^5$ background, where we choose supersymmetric thermal boundary conditions on $\partial Y = S^3 \times S^1_\tau \times S^5$ with supersymmetry preserving twists on $S^1_\tau$. To evaluate the associated string partition function we start from the low-energy effective supergravity action, whose bosonic bulk part equals
\begin{equation}\label{Eq: IIB pseudo action}
\begin{aligned}
	S_{\text{bulk}} =& - \frac{1}{2\kappa^2}\int \left( \star R -\frac12 \rmd \Phi \wedge \star \rmd \Phi - \frac12 \rme^{-\Phi} \rmd B_2 \wedge \star \rmd B_2 \right)\\
	&+\frac{1}{4\kappa^2} \int 	\left( \rme^{2\Phi} F_1 \wedge \star F_1 + \rme^{\Phi}  F_3 \wedge \star  F_3 + \frac12  F_5 \wedge \star F_5 \right)\\
	&+\frac{\rmi}{4\kappa^2} \int C_4 \wedge \rmd B_2 \wedge \rmd C_2\,,
\end{aligned}
\end{equation}
where in string units $\frac{1}{2\kappa^2} = \frac{2\pi}{(2\pi \ell_s)^8}$ and
\begin{equation}
	F_1 = \rmd C_0 \,,\quad F_3 = \rmd C_2 - C_0 \rmd B_2\,,\quad F_5 = \rmd C_4 - \frac12 (C_2 \wedge \rmd B_2 - B_2 \wedge \rmd C_2)\,.
\end{equation}
It is a pseudo action as the self-duality constraint on the five form
\begin{equation}\label{Eq: self duality}
	\star F_5 = +\rmi\, F_5
\end{equation}
is only imposed by hand at the level of the equations of motion.\footnote{\label{foot: orientation}In Euclidean signature $\star^2=-1$ on five-forms, so the eigenvalues of $\star$ are $\pm\rmi$. We fix $\mathrm{vol}_{10}=\mathrm{vol}_{S^5}\wedge\mathrm{vol}_{\text{AdS}_5}$ and positive five-form flux $N>0$ in \eqref{Eq: flux source definitions}. These choices give $\star\mathrm{vol}_{\text{AdS}_5}=-\mathrm{vol}_{S^5}$ and $\star\mathrm{vol}_{S^5}=\mathrm{vol}_{\text{AdS}_5}$, and hence select the $+\rmi$ eigenspace in \eqref{Eq: self duality}.} Instead, one could work with the PST action, which derives the duality constraint directly from varying the action \cite{Pasti:1996vs,DallAgata:1997gnw,DallAgata:1998ahf,Adhikari:2026rfb},
\begin{equation}
	S_{\text{PST}} = S_{\text{bulk}} + \frac{1}{8\kappa^2} \int i_V \mathcal F_5 \wedge \star i_V \mathcal F_5\,,\quad \mathcal F_5 = F_5 + \rmi \star F_5\,,
\end{equation}
where $V = \frac{1}{\sqrt{-|\partial a|^2}} \rmd a$ and $a$ is the standard PST scalar. This action is symmetric under the additional PST gauge transformations
\begin{equation}
\begin{aligned}
	\delta_\eta a =& \eta\,,\quad \delta_\eta C_4 = - \frac{\eta}{\sqrt{-|\partial a|^2}} i_V \mathcal F_5\,,\\
	\delta_b a =& 0\,,\quad \delta_b C_4 = b_4 \,,\quad \text{with} \quad \rmd a \wedge \rmd b_4 = 0\,,
\end{aligned}
\end{equation}
under the constraint that both $\eta$ and $b_4$ vanish on the boundary, fixing Dirichlet boundary conditions for both the PST scalar $a$ and the potential $C_4$. To keep track of which boundary conditions are imposed we split the five-form into an electric and magnetic piece (embedded in AdS$_5$ and $S^5$ respectively) 
\begin{equation}
	F_5 = F_5^{\text{el}} + F_5^{\text{mag}}\,,
\end{equation}
and denote their boundary pullbacks with lower cases. The variation of the PST action reduces, in the self-dual point $\mathcal F_5 = 0$ and at fixed boundary value of $a$, to
\begin{equation}\label{Eq: half variation}
	\delta_{C_4^{\mathrm{el}}} S_{\text{PST}}\Big|
	=\frac{1}{4\kappa^2}\oint_{\partial}
	\delta c_4^{\mathrm{el}}\wedge\star f_5^{\mathrm{el}}
	=\frac12\,\big\langle \delta c_4^{\mathrm{el}},\star f_5^{\mathrm{el}}\big\rangle\,,
	\qquad
	\langle A,B \rangle \equiv \frac{2\pi}{(2\pi \ell_s)^8} \int_\partial A\wedge B\,,
\end{equation}
naturally imposing Dirichlet boundary conditions on the boundary four-form. In the electric/magnetic decomposition the pair of boundary values to keep track of are
\begin{equation}\label{Eq: boundary momentum}
	c_4^{\text{el}}\quad \text{and}\quad \star f_5^{\mathrm{el}}=\rmi\,f_5^{\mathrm{mag}}\,,
\end{equation}
which are conjugate: the electric potential is the source, and the momentum conjugate to it is the magnetic five-form flux. 

The PST equations and gauge fixing gives $\mathcal F_5=0$, upholding the self-duality constraint of type IIB \eqref{Eq: self duality}. Although the PST action is fully consistent with the IIB equations of motion, it is not by itself the physical action since its self-dual on-shell value vanishes, and it in fact does not provide a consistent variational principle. The fact that the on-shell action of IIB vanishes in AdS$_5 \times S^5$ is obvious from
\begin{equation}\label{Eq: F5 illustration}
	R = 0\,,\qquad F_5 = -\frac{4}{L}\Big(\rmi\,\mathrm{vol}_{\text{AdS}_5} + \star\,\mathrm{vol}_{\text{AdS}_5}\Big) = -\frac{4\rmi}{L}\,\mathrm{vol}_{\text{AdS}_5}+\frac{4}{L}\,\mathrm{vol}_{S^5}\,,
\end{equation}
and
\begin{equation}
	\star F_5=+\rmi F_5\,,\qquad \mathcal F_5=F_5+\rmi\star F_5=0\,,
\end{equation}
making the action integrand vanish identically. The proposed resolution of \cite{Kurlyand:2022vzv} adds a topological term to the action of the form\footnote{In \cite{Adhikari:2026rfb} this analysis was expanded on to backgrounds with additional non-trivial fluxes.}
\begin{equation}\label{Eq: topological term}
	S_{\text{top}} = -\frac{\rmi}{4\kappa^2}
	\int F_5^{\mathrm{el}}\wedge F_5^{\mathrm{mag}}\,,
\end{equation}
a total derivative, which changes no equation of motion but does change boundary terms and the on-shell value of the action. To fix the coupling we will add this topological piece to the action with an arbitrary constant\footnote{We added the GHY term to fix Dirichlet boundary conditions on the metric, and $S_{\text{ct}}$ contains any possible additional boundary terms needed to regulate the action.}
\begin{equation}\label{Eq: c family}
	S^{(\alpha)}_{\text{IIB}} = S_{\text{PST}} + \alpha\,S_{\text{top}} + S_{\text{GHY}} + S_{\text{ct}}\,.
\end{equation}
We now repeat the PST boundary calculation of \cite{Kurlyand:2022vzv} for arbitrary coupling, holding nothing fixed for the moment.  Because $F_5^{\mathrm{el}}=\rmd C_4^{\mathrm{el}}$ and $\rmd F_5^{\mathrm{mag}}=0$, the topological term is a boundary pairing of exactly the two data of \eqref{Eq: boundary momentum},
\begin{equation}\label{Eq: topological boundary form}
	S_{\mathrm{top}}
	=-\frac{\rmi}{4\kappa^2}\int F_5^{\mathrm{el}}\wedge F_5^{\mathrm{mag}}
	=-\frac{\rmi}{4\kappa^2}\oint_{\partial} c_4^{\mathrm{el}}\wedge f_5^{\mathrm{mag}}
	=-\frac12\,\big\langle c_4^{\mathrm{el}},\star f_5^{\mathrm{el}}\big\rangle\,,
\end{equation}
so that its variation equals
\begin{equation}\label{Eq: topological variation}
	\delta S_{\mathrm{top}}
	=-\frac12\,\big\langle\delta c_4^{\mathrm{el}},\star f_5^{\mathrm{el}}\big\rangle
	-\frac12\,\big\langle c_4^{\mathrm{el}},\delta(\star f_5^{\mathrm{el}})\big\rangle\,.
\end{equation}
The second term is not to be varied continuously, $\star f_5^{\mathrm{el}}$ is a quantised closed form and labels a flux sector. A variational problem is to be solved within one such sector, $\delta(\star f_5^{\mathrm{el}})=0$, and one is either in a fixed flux sector or sums over them.  Within a single sector, \eqref{Eq: half variation} and \eqref{Eq: topological variation} combine so that the boundary variation equals
\begin{equation}\label{Eq: generic c variation}
	\delta S^{(\alpha)}_{\mathrm{IIB}}
	=\frac{1-\alpha}{2}\,\big\langle\delta c_4^{\mathrm{el}},\star f_5^{\mathrm{el}}\big\rangle\,.
\end{equation}
Now it is clear that the choice of coupling for the topological term is closely related to the choice of ensemble. Labelling the flux number and conjugate chemical potential with
\begin{equation}\label{Eq: flux source definitions}
	N=\frac{1}{(2\pi\ell_s)^4}\int_{S^5}F_5^{\mathrm{mag}}\,,
	\qquad
	\mu=\frac{2\pi\rmi}{(2\pi\ell_s)^4}
	\int_{S^1_\tau\times S^3}c_4^{\mathrm{el}}\,,
\end{equation}
the pairing of \eqref{Eq: topological boundary form} reads $\big\langle c_4^{\mathrm{el}},\star f_5^{\mathrm{el}}\big\rangle=-\mu N$, so that \eqref{Eq: generic c variation} becomes
\begin{equation}\label{Eq: generic c variation in mu}
	\delta S^{(\alpha)}_{\mathrm{IIB}}
	=-\frac{1-\alpha}{2}\,N\,\delta\mu\,.
\end{equation}
If the magnetic flux is held fixed, the electric potential must be free to fluctuate at the boundary and \eqref{Eq: generic c variation in mu} has to vanish for arbitrary $\delta\mu$. This happens only at $\alpha=+1$, which is the coupling fixed in \cite{Kurlyand:2022vzv}. If instead the electric potential is held fixed on the boundary, $\delta c_4^{\mathrm{el}}=0$ annihilates \eqref{Eq: generic c variation} for every $\alpha$ and the action is stationary regardless. What fixes $\alpha$ here is then not stationarity but the normalisation of the response. The source is periodic, $\mu\sim\mu+2\pi\rmi$ by the large gauge transformations of $c_4^{\mathrm{el}}$ \eqref{Eq: large gauge}, so the charge conjugate to it is quantised and since $\star f_5^{\mathrm{el}}$ is the only response available, that charge must be the quantised flux $N$ itself rather than a multiple or a fraction of it. In other words, to fix the coefficient $\alpha$ in the $\mu$ ensemble we impose the Hamilton--Jacobi relation
\begin{equation}\label{Eq: HJ fixes alpha}
	\frac{\partial S^{(\alpha)}_{\text{IIB}}}{\partial\mu}
	=-\frac{1-\alpha}{2}\,N
	\;=\;-N
	\qquad\Rightarrow\qquad
	\alpha=-1\,.
\end{equation}
To summarise, we have $(c_4^{\mathrm{el}},\star f_5^{\mathrm{el}})$ as a conjugate pair of which only one can be held fixed on the boundary, and $\alpha$ is nothing but the label of which one that is. The fixed flux ensemble has $\alpha=+1$ by stationarity, while the fixed potential ensemble has $\alpha=-1$ by the normalisation of the response.  The two are boundary choices which fix different polarizations within the same theory and are related by a standard Legendre transform. With the couplings of the topological terms fixed in this way the Hamilton--Jacobi relations read
\begin{equation}\label{Eq: HJ check}
	\frac{\partial S^{(\mu)}_{\text{IIB}}}{\partial\mu}=-N\,,
	\qquad \text{and}\qquad 
	\frac{\partial S^{(N)}_{\text{IIB}}}{\partial N}=+\mu\,.
\end{equation}
The variational principle therefore admits two consistent ensembles
\begin{equation}\label{Eq: two actions}
	S^{(\mu)}_{\text{IIB}} = S_{\text{PST}} - S_{\text{top}} + S_{\text{GHY}} + S_{\text{ct}}\,,
	\qquad
	S^{(N)}_{\text{IIB}} = S_{\text{PST}} + S_{\text{top}} + S_{\text{GHY}} + S_{\text{ct}}\,,
\end{equation}
fixed by the sign of the topological term. The two choices are conjugate to one another, describe the same theory but with different boundary polarizations.  Their difference is twice the topological term, which by \eqref{Eq: topological boundary form} and \eqref{Eq: flux source definitions} is the Legendre transform
\begin{equation}\label{Eq: boundary variation}
	\rmd S^{(N)}_{\text{IIB}}=\rmd S^{(\mu)}_{\text{IIB}}+N\,\rmd\mu+\mu\,\rmd N=\mu\,\rmd N\,.
\end{equation}
On the boundary the same relation is the statement that the two partition functions are related by $\widehat\Xi(\mu)=\sum_N\rme^{\mu N} Z_N$, whose semi-classical evaluation is the Legendre transform of the fixed-rank answer. The point is that in the bulk, using \eqref{Eq: two actions}, both the canonical and grand canonical ensembles can be studied. 

Finally, the relevant PST gauge group is polarization dependent. Under $\delta_bC_4=b_4$, we find that
\begin{equation}\label{Eq: PST boundary gauge variation}
	\delta_b S_{\text{PST}}
	=+\frac{\rmi}{4\kappa^2}\oint_\partial F_5^{\mathrm{mag}}\wedge b_4\,,
	\qquad
	\delta_b S_{\text{top}}
	=-\frac{\rmi \alpha}{4\kappa^2}\oint_\partial F_5^{\mathrm{mag}}\wedge b_4\,,
\end{equation}
which cancel each other exactly for $\alpha = +1$. For $\alpha=-1$ we have to impose instead that $b_4|_\partial=0$, which is in fact entirely consistent with fixing Dirichlet boundary conditions for $C_4^{\text{el}}$.

The choice of holographic ensemble in IIB is thus more subtle than in M-theory, where the natural ensemble was imposed because the standard local supergravity action in eleven-dimensions is written in terms of $A_3$, not $A_6$ \cite{Gautason:2025plx,vanMuiden:2026nsp,Bobev:2026gir}. In type IIB string theory there does not seem to be a natural ensemble due to the self-dual nature of the five-form and the PST formulation. Working either with fixed flux or fixed chemical potential is fully determined by the sign of the added topological term.

The following sections study the bulk dynamics in both the fixed-$c_4^{\text{el}}$ and fixed-$\star f_5^{\text{el}}$ ensembles and match them to the boundary observables in the grand-canonical and canonical ensembles, respectively.
\subsection{Supersymmetric thermal AdS}\label{sec: schur background}
The IIB backgrounds dual to the supersymmetric $S^3\times S^1$ partition functions of the boundary $\mathcal N=4$ theory are quotients of Euclidean global AdS$_5\times S^5$. Compactifying Euclidean time alone breaks every supersymmetry since the Killing spinors of AdS$_5\times S^5$ depend explicitly on $\tau$, and the remedy is to twist the thermal circle by the angular isometries of the geometry. On the three-sphere inside AdS$_5$ two commuting rotations are available, and on the internal $S^5$ three more. Explicitly writing the metric with all twists we have that
\begin{equation}\label{Eq: Schur background}
\begin{aligned}
	\rmd s^2_{10} &= L^2\Big[\rmd\rho^2 + \cosh^2\!\rho\,\rmd \tau^2 + \sinh^2\!\rho\,\rmd\Omega_3^2\Big] + L^2 \rmd \Omega_5^2 \,,\\
	\rmd\Omega_3^2 &= \rmd\xi^2 + \sin^2\!\xi\,\big(\rmd\varphi_1+\rmi\,\varepsilon_1\,\rmd \tau\big)^2 + \cos^2\!\xi\,\big(\rmd\varphi_2 + \rmi\,\varepsilon_2\,\rmd \tau\big)^2\,,\\
	\rmd \Omega_5^2 &= \sum_{I=1}^{3}\Big[\rmd\mu_I^2+\mu_I^2\big(\rmd\phi_I+\rmi\,\sigma_I\,\rmd\tau\big)^2\Big]\,,\qquad \sum_{I=1}^{3}\mu_I^2=1\,,
\end{aligned}
\end{equation}
where we use the parametrisation $(\mu_1,\mu_2,\mu_3)=(\cos\theta,\sin\theta\sin\chi,\sin\theta\cos\chi)$ on the five-sphere. The twists in AdS correspond to rotational data on the boundary and the twists on $S^5$ to R-symmetry data. Both spheres are round in the shifted angles
\begin{equation}\label{Eq: twisted angles}
	\tilde\varphi_a = \varphi_a+\rmi\,\varepsilon_a\,\tau\,,
	\qquad
	\tilde\phi_I = \phi_I+\rmi\,\sigma_I\,\tau\,,
\end{equation}
in terms of which $\rmd\Omega_5^2=\rmd\theta^2+\cos^2\!\theta\,\rmd\tilde\phi_1^2+\sin^2\!\theta\,\rmd\hat\Omega_3^2$, with $\hat S^3$ the unit three-sphere spanned by $(\chi,\tilde\phi_2,\tilde\phi_3)$. The background is supported by the four-form potential
\begin{equation}\label{Eq: C4 background}
\begin{aligned}
	C_4^{\mathrm{el}}
	&=-\rmi L^4\sinh^4\!\rho\;\rmd\tau\wedge
	\mathrm{vol}_{\tilde S^3}\,,\\
	C_4^{\mathrm{mag}}
	&=L^4\sin^4\!\theta\;\rmd\tilde\phi_1\wedge\mathrm{vol}_{\hat S^3}\,,
	\qquad C_4=C_4^{\mathrm{el}}+C_4^{\mathrm{mag}}\,,
\end{aligned}
\end{equation}
whose exterior derivative is the five form
\begin{equation}\label{Eq: F5 background}
	F_5^{\mathrm{el}}=-\frac{4\rmi}{L}\,\mathrm{vol}_{\text{AdS}_5}\,,
	\qquad
	F_5^{\mathrm{mag}}=\frac{4}{L}\,\mathrm{vol}_{S^5}\,,
\end{equation}
with the volume forms of the twisted metric \eqref{Eq: Schur background}. The two gauge potentials treat the twists differently. In $C_4^{\mathrm{el}}$ the $\rmi\,\varepsilon_a\,\rmd\tau$ legs of $\mathrm{vol}_{\tilde S^3}$ drop against the overall $\rmd\tau$, so the electric potential is insensitive to the twists; in $C_4^{\mathrm{mag}}$ there is no such $\rmd\tau$ and the $\rmi\,\sigma_I\,\rmd\tau$ legs survive. The flux number and chemical potential in this background equal
\begin{equation}\label{Eq: flux and chemical potential}
\begin{aligned}
	N
	&=\frac{1}{(2\pi\ell_s)^4}\int_{S^5}F_5^{\mathrm{mag}}
	=\frac{L^4}{4\pi\ell_s^4}\,,\\
	\mu
	&=\frac{2\pi\rmi}{(2\pi\ell_s)^4}
	\int_{S^1_\tau\times S^3}c_4^{\mathrm{el}}
	=\frac{L^4}{2\pi^3\ell_s^4}\,
	\text{Vol}^{\text{ren}}_{\text{AdS}_5}\,.
\end{aligned}
\end{equation}
We leave the renormalised volume undetermined for now as its finite value will depend on the supersymmetric scheme, and the background twist parameters.

These twists are simple coordinate transformations and do not alter the equations of motion, but instead they globally identify the periods such that
\begin{equation}\label{Eq: twisted identification}
	(\tau,\,\tilde\varphi_a,\,\tilde\phi_I)\ \sim\ (\tau+\beta,\ \tilde\varphi_a+\rmi\,\varepsilon_a\beta,\ \tilde\phi_I+\rmi\,\sigma_I\beta)\,,
\end{equation}
in the shifted angles \eqref{Eq: twisted angles}. Supersymmetry is then a condition on this identification. The Killing spinors of AdS$_5\times S^5$ are known in closed form (see e.g. appendix A of \cite{Gautason:2024nru}) and we can organise them into simultaneous eigenspinors along the five twisted Killing vectors. Under the transformation in \eqref{Eq: twisted identification} an eigenspinor is multiplied by $\rme^{\beta(\Delta-\varepsilon_1 j_1-\varepsilon_2 j_2-\sum_I\sigma_I r_I)}$ and to ensure the spinors are periodic and we preserve supersymmetry we must impose
\begin{equation}\label{Eq: susy condition}
	\Delta = \varepsilon_1\, j_1+\varepsilon_2\, j_2+\sum_{I=1}^{3}\sigma_I\, r_I\,,
\end{equation}
which for the weights $\pm(\tfrac12,\tfrac12,\tfrac12,\tfrac12,\tfrac12,\tfrac12)$ fixes the supersymmetric family
\begin{equation}\label{Eq: susy twist family}
	\varepsilon_1+\varepsilon_2+\sigma_1+\sigma_2+\sigma_3 = 1\,.
\end{equation}
These are all the supersymmetry preserving thermal boundary conditions available on $S^3\times S^1$ with the round metric untouched. In the index variables $\omega_a=\beta(1+\varepsilon_a)$ and $\Delta_I=\beta(1-\sigma_I)$ the family \eqref{Eq: susy twist family} is the statement $\sum_I\Delta_I=\sum_a\omega_a$, that is the $n=0$ branch of \eqref{Eq: appendix CCMM}. Of the thirty-two Killing spinors of AdS$_5\times S^5$ two survive at a generic point of \eqref{Eq: susy twist family}, which is enhanced by restricting to the Schur limit taking
\begin{equation}\label{Eq: Schur twist}
	(\varepsilon_1,\varepsilon_2) = (1,0)\,,\qquad \sigma_1=\sigma_2=\sigma_3=0\,,
\end{equation}
for which the condition \eqref{Eq: susy condition} degenerates to $\Delta=j_1$, $j_2$ is left unconstrained. The ten-dimensional Killing spinor is made $\tau$-independent through the projector $\tfrac12(1+\rmi\Gamma_{12})\epsilon_0=0$ \cite{Gautason:2024nru}. A useful check of \eqref{Eq: susy condition} at the opposite corner of the family is the purely internal twist $\varepsilon_a=0$, $\sigma_I=(1,0,0)$, the half-BPS twist of \cite{Gautason:2024nru}: there \eqref{Eq: susy condition} reads $\Delta=r_1$ and the projector to make the spinor $\tau$-independent is $\tfrac12(1-\rmi\Gamma_{5\,10})\epsilon_0=0$.

Since these twists are coordinate transformations, they are invisible to any local quantity in the bulk, but they do change the complex structure. This will become important when we compute the on-shell action, so let us be more explicit. In the untwisted coordinates the boundary of the AdS geometry is parametrised by
\begin{equation}\label{Eq: Hopf coordinates}
	z_1 = \rme^{-\tau+\rmi\tilde\varphi_1}\,\sin\xi\,,\qquad
	z_2 = \rme^{-\tau+\rmi\tilde\varphi_2}\,\cos\xi\,,
\end{equation}
on which the identification \eqref{Eq: twisted identification} acts holomorphically,
\begin{equation}\label{Eq: Hopf moduli}
	(z_1,z_2)\ \sim\ \big(\rme^{-\beta b_1}\,z_1,\ \rme^{-\beta b_2}\,z_2\big)\,,\qquad
	b_a = 1+\varepsilon_a\,.
\end{equation}
The quotient is determined by the complex structure moduli $(\beta b_1,\beta b_2)$ \cite{Assel:2014paa}, and for the moduli on the supersymmetric family \eqref{Eq: susy twist family} we have for $\sum_I \sigma_I = 0$ that
\begin{equation}\label{Eq: susy moduli slice}
	b_1+b_2 = 3\,,
\end{equation}
which further get fixed for the Schur twist \eqref{Eq: Schur twist} where
\begin{equation}\label{Eq: Schur moduli}
	(b_1,b_2) = (2,1)\,. 
\end{equation}
These global aspects will be important when choosing finite supersymmetric counterterms to regulate the AdS volume below. They do not change the local bulk integrand, however, and in the $\mu$ ensemble the action evaluates to
\begin{equation}
	S^{(\mu)}_{\text{IIB}} = - \frac{L^8}{16 \pi^4 \ell_s^8} \text{Vol}_{\text{AdS}_5}^{\text{ren}}\,.
\end{equation}
Replacing the length scales with the chemical potential using \eqref{Eq: flux and chemical potential} we find that
\begin{equation}
	S^{(\mu)}_{\text{IIB}} = -\frac{\pi^2}{4\,\text{Vol}_{\text{AdS}_5}^{\text{ren}}}\,\mu^2\,.
\end{equation}
What remains is the renormalised volume of AdS. In the minimal scheme, with only local geometric boundary terms to remove the divergences \cite{Henningson:1998gx,Emparan:1999pm,Skenderis:2002wp} its volume equals
\begin{equation}\label{Eq: minimal volume}
	\text{Vol}^{\text{ren}}_{\text{AdS}_{2p+1}} = \left(\frac{-\pi}{4}\right)^{p} \frac{\Gamma(2p+1)}{\Gamma(p+1)^3}\,\beta
	\qquad \Rightarrow \qquad
	\text{Vol}^{\text{ren},\,\text{min}}_{\text{AdS}_5} = \frac{3\pi^2}{16}\,\beta\,.
\end{equation}
In this scheme the on-shell action equals
\begin{equation}
	S^{(\mu)}_{\text{IIB}} = -\frac{4}{3\beta}\,\mu^2\,.
\end{equation}
This answer is independent of the moduli $(b_1,b_2)$ while the boundary partition function depends explicitly on the twists through those moduli \cite{Assel:2014paa,Assel:2015nca}. Inspired by this mismatch it was argued in  \cite{BenettiGenolini:2016qwm,BenettiGenolini:2016tsn} that the minimal scheme is in fact not supersymmetric. The supersymmetric scheme instead contains non-local finite counter terms, which are uniquely determined by the complex structure moduli. Including these counterterms the renormalized volume equals\footnote{In \cite{BenettiGenolini:2016qwm,BenettiGenolini:2016tsn} the authors used these counterterms to regularize the on-shell action of five-dimensional minimal gauged supergravity, which is a consistent truncation of type IIB on $S^5$. Here we only need the relevant results to renormalize Euclidean global AdS$_5$.}
\begin{equation}
	\text{Vol}^{\text{ren},\,\text{susy}}_{\text{AdS}_5}
	= \frac{\pi^2}{54}\,\frac{(b_1+b_2)^3}{b_1b_2}\,\beta\,, 
\end{equation}
such that
\begin{equation}
	S_{\text{IIB}}^{(\mu)} =\, -\frac{27 b_1 b_2}{2 (b_1 + b_2)^3} \frac{\mu^2}{\beta}\qquad 
		\mu =\, \frac{(b_1+b_2)^3\beta}{108 \pi b_1 b_2} \frac{L^4}{\ell_s^4}\,,
\end{equation}
and in the Schur limit 
\begin{equation}
	S_{\text{IIB}}^{(\mu)} = -\frac{\mu^2}{\beta} \,,\qquad \mu = \frac{\beta}{8 \pi} \frac{L^4}{\ell_s^4}\,.
\end{equation}
It is this on-shell action which we will match to the large $\mu$ limit of the grand canonical partition function of the boundary theory in the next section. To instead go to the fixed flux ensemble, in the large $L/\ell_s$ limit we perform the Legendre transform and find
\begin{equation}\label{Eq: leading sugra answers}
	S^{(\mu)}_{\text{IIB}}=-\frac{\mu^2}{\beta}\,,
	\qquad
	S^{(N)}_{\text{IIB}}
	=\left[S^{(\mu)}_{\text{IIB}}+\mu N\right]_{\mu=\mu_\ast}
	=\frac{\beta N^2}{4}\,
\end{equation}
with the classical saddle at $\mu_*=\beta N/2$. In the fixed-flux ensemble the on-shell action corresponds to the supersymmetric Casimir energy, which we review in appendix~\ref{App: casimir}.

The classical action fixes the leading quadratic $\mu$-dependence. In general one is to expect additional $1/\mu$ corrections in the full partition function, with the first one being of the order $\log \mu$ due to zero-modes running in a loop. In the twisted Schur background there are no such normalisable modes present and thus there is no such log correction. 

The subsequent correction is of order $\mu^0$ and arises from the massive modes running in the loop.  We do not evaluate their determinant directly but instead use the single-particle supergravity index of \cite{Kinney:2005ej}, whose plethystic exponential is the corresponding one-loop index partition function, setting the fugacities of \cite{Kinney:2005ej} to $(t,y,v,w) = (q^{2/3}, 1, q^{-1/3},q^{1/3})$ (with $q=\rme^{-\beta}$) to specialise to the Schur limit, such that
\begin{equation}\label{Eq: Schur KK single particle}
	i_{\text{KK}}(\beta)= \frac{2 + \rme^{-\beta}}{2\sinh\beta}\,,
\end{equation}
where we have taken both the local supergravity fields and the singleton multiplet into account since we are aiming to reproduce the boundary partition function with $\text{U}(N)$ gauge group. The plethystic exponential of the single-particle index equals
\begin{equation}\label{Eq: KK determinant}
	\Akk(\beta) = \prod_{n=1}^{\infty} \frac{1}{(1-q^{2n})(1-q^{2n-1})^2} = \frac{1}{\vartheta_4(0|\tau)}\,,
\end{equation}
so that a single bulk saddle contributes
\begin{equation}\label{Eq: single saddle}
	-S^{(\mu)}_{\text{IIB}}+\log\Akk(\beta)
	= \frac{\mu^2}{\beta} + \log \Akk(\beta)\,.
\end{equation}
We have not derived that there are no further $1/\mu$ corrections, but we will see upon comparison to the boundary partition function that those must in fact vanish. Proving this statement directly in the bulk would seem to require a direct supersymmetric localization procedure of the IIB theory. As already emphasised, the boundary value of $c_4^{\mathrm{el}}$ is defined only up to large gauge transformations,
\begin{equation}\label{Eq: large gauge}
	\frac{2\pi\rmi}{(2\pi\ell_s)^4}
	\int_{S^1_\tau\times S^3}c_4^{\mathrm{el}}
	\longrightarrow
	\frac{2\pi\rmi}{(2\pi\ell_s)^4}
	\int_{S^1_\tau\times S^3}c_4^{\mathrm{el}}+2\pi\rmi\,,
\end{equation}
or equivalently $\mu\sim\mu+2\pi\rmi$. The full partition function is subsequently invariant under these large gauge transformations, which can only be the case if we sum over all large gauge images of \eqref{Eq: single saddle}. Combining all ingredients, and introducing an overall normalisation, which can be attributed to the string measure, we find
\begin{equation}\label{Eq: IIB pert part func in mu}
	Z_{\text{IIB}}^{\text{pert}}(\mu,\beta) = \sum_{n\in\Z}
	\exp\!\left[\mathcal Z_{\text{IIB}}^{\text{pert}}(\mu + 2\pi \rmi n,\beta)\right] = \mathcal N \, \Akk(\beta)  \sum_{n\in\Z}
	\exp\!\left[\frac{(\mu+2\pi\rmi n)^2}{\beta}\right]\,,
\end{equation}
where $Z_{\text{IIB}}^{\text{pert}}(\mu,\beta) $ denotes the perturbative part of the IIB partition function including the sum over large gauge images and
\begin{equation}
	\mathcal Z_{\text{IIB}}^{\text{pert}}(\mu,\beta) = \frac{\mu^2}{\beta}
	+ \log\!\left[\mathcal N\,\mathcal A_{\text{KK}}(\beta)\right]\,.
\end{equation}
Deriving the overall normalisation from first principles would have to come from a quantisation of the IIB string. We instead fix this number by imposing that the dual conjugate partition function trivialises in the zero flux sector
\begin{equation}\label{Eq: ensemble measure}
	\frac{1}{2\pi\rmi}\int_{-\rmi\pi}^{\rmi\pi}\!\rmd\mu\;
	\mathcal N\sum_{n\in\Z}\exp\!\left[\frac{(\mu+2\pi\rmi n)^2}{\beta}\right]=1
	\quad\Rightarrow\quad
	\mathcal N=\sqrt{\frac{4\pi}{\beta}}\,.
\end{equation}
With this value the sum over large gauge images is precisely a Jacobi theta function with $\mathcal N=\sqrt{4\pi/\beta}$ being its automorphy factor upon an S-transform on $\rmi\beta/4\pi$, and \eqref{Eq: IIB pert part func in mu} collapses to\footnote{We defined the $\vartheta$-functions in \eqref{Eq: theta conventions}.}
\begin{equation}\label{Eq: IIB theta form}
	Z^{\text{pert}}_{\text{IIB}}(\mu,\beta)
	=\mathcal N\,\Akk(\beta)\sum_{n\in\Z}
	\exp\!\left[\frac{(\mu+2\pi\rmi n)^2}{\beta}\right]
	=\Akk(\beta)\,\vartheta_3\!\left(\frac{\rmi\mu}{2}\Big|\frac{\rmi\beta}{4\pi}\right)\,.
\end{equation}
For non-trivial values of the flux we subsequently find that the partition function equals
\begin{equation}\label{Eq: IIB fixed flux}
\begin{aligned}
	Z^{\text{pert}}_{\text{IIB}}(N,\beta)
	=&\,\frac{1}{2\pi\rmi} \int_{-\rmi \pi}^{\rmi \pi} \rmd \mu \;\rme^{-N\mu}\,
	\Akk(\beta) \sqrt{\frac{4\pi}{\beta}} \sum_{n\in\Z}
	\exp\!\left[\frac{(\mu+2\pi\rmi n)^2}{\beta}\right]\\
	=&\,\Akk(\beta)\sqrt{\frac{4\pi}{\beta}}\;\frac{1}{2\pi\rmi}
	\int_{-\rmi\infty}^{+\rmi\infty}
	\rmd\mu\;\exp\!\left(\frac{\mu^2}{\beta}-N\mu\right)
	= \Akk(\beta)\,\rme^{-\frac{\beta N^2}{4}}\,,
\end{aligned}
\end{equation}
where $Z^{\text{pert}}_{\text{IIB}}(N,\beta)$ is defined at fixed flux and does not require a sum over large gauge images.

In the next section we will review the boundary Schur partition function in the canonical and grand canonical ensemble and show that they equal \eqref{Eq: IIB fixed flux} and \eqref{Eq: IIB pert part func in mu} respectively.

For boundaries of more complicated topology the measure and contour of the ensemble changing integral are an open question, analysed for AdS$_4$ in \cite{Bobev:2026gir}, and would be interesting to understand in more detail. This would give access to the partition function in both ensembles for more complicated backgrounds including the black hole and black string backgrounds.
\section{The boundary grand canonical ensemble}\label{sec: grand canonical}
We now turn to the boundary side of the computation \cite{Bourdier:2015wda} and show that the results in that paper are in perfect agreement with our analysis here, in both ensembles. Supersymmetric localisation gives the following matrix-model representation \cite{Romelsberger:2005eg,Kinney:2005ej,Gadde:2011ik,Gadde:2011uv} of the fixed-rank partition function on $S^1\times S^3$:
\begin{equation}\label{Eq: physical Schur partition function}
	\begin{aligned}
	Z_N(\beta)
	\equiv q^{N^2/4}\mathcal I_N(q) 
	=\frac{\eta(\tau)^{3N}}{N!\,\pi^N}
	\int_{[0,\pi]^N}\!\rmd^N\alpha\,
	\frac{\displaystyle\prod_{i<j}
	\vartheta_1(\alpha_i-\alpha_j|\tau)^2}
	{\displaystyle\prod_{i,j}
	\vartheta_4(\alpha_i-\alpha_j|\tau)}\,.
	\end{aligned}
\end{equation}
Here $\mathcal I_N(q)$ is the Schur index of $\mathcal N=4$ $\UU(N)$ SYM.  The factor $q^{N^2/4} =\rme^{-\beta E_{\rm susy}(N)}$, with $E_{\rm susy}(N)=N^2/4$, relates the index to the supersymmetric partition function.\footnote{Appendix~\ref{App: casimir} provides an independent derivation of this factor.}  In string theory we aim to match bulk and boundary partition functions, and thus it is $Z_N$, rather than $\mathcal I_N$, which is our observable of interest. 

The Jacobi theta functions appearing in \eqref{Eq: physical Schur partition function} are defined as
\begin{equation}\label{Eq: theta conventions}
	\begin{aligned}
	\vartheta_1(z|\tau)
	&=-\rmi\sum_{n\in\Z}(-1)^n q^{(n+\frac12)^2}
	  \rme^{\rmi(2n+1)z}\,,
	&\vartheta_2(z|\tau)
	&=\sum_{n\in\Z}q^{(n+\frac12)^2}\rme^{\rmi(2n+1)z}\,,\\
	\vartheta_3(z|\tau)
	&=\sum_{n\in\Z}q^{n^2}\rme^{2\rmi nz}\,,
	&\vartheta_4(z|\tau)
	&=\sum_{n\in\Z}(-1)^n q^{n^2}\rme^{2\rmi nz}\,,
	\end{aligned}
\end{equation}
and
\begin{equation}\label{Eq: eta convention}
	q=\rme^{\rmi\pi\tau}=\rme^{-\beta} \,,\qquad
	\eta(\tau)=q^{1/12}\prod_{m=1}^{\infty}(1-q^{2m})\,.
\end{equation}
In \cite{Bourdier:2015wda} it was shown that the grand canonical partition function takes a particularly simple form
\begin{equation}\label{Eq: exact BDF grand index}
\begin{aligned}
	\Xie(\mu,\tau)
	&\equiv\sum_{N=0}^{\infty} Z_N(\beta)\rme^{\mu N}
	=\frac{\vartheta_3(\rmi \mu_{\text{eff}}|\tau)+\vartheta_2(\rmi \mu_{\text{eff}}|\tau)}
	{\vartheta_4(0|\tau)}\,,\\
	\mu_{\text{eff}}
	&=\mu+\log\!\left[
	\frac{1+\sqrt{1-4\rme^{-2\mu}}}{2}\right] \,.
\end{aligned}
\end{equation}
This effective chemical potential gets non-perturbatively renormalised at large values, i.e.
\begin{equation}\label{Eq: approximate W}
	\mu_{\text{eff}} = \mu + \mathcal O(\rme^{-2\mu}).
\end{equation}
Excluding these non-perturbative contributions for now the grand canonical partition function simplifies to
\begin{align}\label{Eq: perturbative equivalent forms}
	\Xip(\mu,\tau)
	&=\frac{\vartheta_2(\rmi\mu|\tau)+\vartheta_3(\rmi\mu|\tau)}
	{\vartheta_4(0|\tau)}
	=\frac{\vartheta_3\!\bigl(\frac{\rmi\mu}{2}\big|\frac{\tau}{4}\bigr)}
	{\vartheta_4(0|\tau)}\nonumber\\
	&=\Aa(\tau)\exp\!\left(\frac{\rmi\mu^2}{\pi\tau}\right)
	\vartheta_3\!\left(\frac{2\rmi\mu}{\tau}\Big|-\frac{4}{\tau}\right)\nonumber\\
	&=\Aa(\tau)\sum_{n\in\Z}
	\exp\!\left[\frac{\rmi(\mu+2\pi\rmi n)^2}{\pi\tau}\right]
	=\sum_{n\in\Z}\exp\!\left[
	\Jp(\mu+2\pi\rmi n,\tau)\right] \,,
\end{align}
where we applied a modular transform in the second line. We have separated the multiplicative amplitude from its additive contribution to the grand potential by defining
\begin{equation}\label{Eq: boundary perturbative grand potential}
	\Aa(\tau)=\frac{2}{\sqrt{-\rmi\tau}\,\vartheta_4(0|\tau)}\,,
	\qquad
	\Jp(\mu,\tau)=\frac{\rmi\mu^2}{\pi\tau}+\log\Aa(\tau)\,.
\end{equation}
Since $\sqrt{-\rmi\tau}=\sqrt{\beta/\pi}$, the constant contribution is
\begin{equation}\label{Eq: amplitude identification}
	\Aa(\tau)=\frac{2}{\sqrt{-\rmi\tau}\,\vartheta_4(0|\tau)}
	=\sqrt{\frac{4\pi}{\beta}}\,\Akk(\beta)\,,
\end{equation}
matching \eqref{Eq: IIB pert part func in mu}. The perturbative expressions therefore agree exactly
\begin{equation}\label{Eq: perturbative bulk boundary match}
	\Jp(\mu,\beta) = \mathcal Z_{\text{IIB}}^{\text{pert}}(\mu,\beta)\,,
	\qquad
	\Xip(\mu,\beta)=Z^{\text{pert}}_{\text{IIB}}(\mu,\beta)\,.
\end{equation}
The agreement holds in the canonical ensemble as well. Collecting $\vartheta_2$ and $\vartheta_3$ into a single sum over flux sectors
\begin{equation}\label{Eq: perturbative signed flux sum}
	\Xip(\mu,\beta)
	=\frac{1}{\vartheta_4(0|\tau)}\sum_{N\in\Z}q^{N^2/4}\,\rme^{N\mu}
	=\sum_{N\in\Z}\rme^{\mu N}\,\Akk(\beta)\,\rme^{-\frac{\beta N^2}{4}}\,.
\end{equation}
The coefficient of $\rme^{\mu N}$ is therefore
\begin{equation}\label{Eq: fixed N match}
	 Z^{\text{pert}}(N,\beta)
	=\Akk(\beta)\,\rme^{-\frac{\beta N^2}{4}}
	=Z^{\text{pert}}_{\text{IIB},N}(N,\beta)\,,
\end{equation}
which is \eqref{Eq: IIB fixed flux} on the nose and the exponent is the supersymmetric Casimir contribution, reviewed in appendix~\ref{App: casimir}, matching the bulk on-shell action at fixed flux \eqref{Eq: leading sugra answers}. 

The perturbative bulk saddle does not reproduce the information discarded by the approximation of $\mu_{\text{eff}}$ in \eqref{Eq: approximate W}.  The resulting difference is exponentially suppressed
\begin{equation}\label{Eq: correction law}
	\frac{\Xie(\mu,\tau)}{\Xip(\mu,\tau)}-1
	=-\frac{2\mu}{\beta}\,\rme^{-2\mu}
	+\ldots 
\end{equation}
at large $\mu$. The exact grand canonical partition function equals \cite{Bourdier:2015wda}
\begin{equation}\label{Eq: master identity}
	\Xie(\mu,\tau)
	=\frac{\vartheta_3(\rmi\mueff|\tau)+\vartheta_2(\rmi\mueff|\tau)}
	{\vartheta_4(0|\tau)}
	=\Xip\bigl(\mueff(\mu),\tau\bigr)\,,
\end{equation}
and the effective chemical potential is also periodic 
\begin{equation}
	\mu_{\text{eff}}(\mu + 2\pi \rmi n) = \mu_{\text{eff}} + 2\pi\rmi n
\end{equation}
such that the partition function can once again be written as
\begin{align}
	\Xie(\mu,\tau) = \sum_{n\in\Z}\exp\!\left[
	\Je(\mu+2\pi\rmi n,\tau)\right] = \sum_{n\in\Z}\exp\!\left[
	\Jp\bigl(\mueff(\mu)+2\pi\rmi n,\tau\bigr)\right]\nonumber\, .
\end{align}
Since \eqref{Eq: master identity} says that the exact grand potential is the perturbative one evaluated on $\mueff$, the whole non-perturbative sector is the difference of two Gaussians, $J_{\text{np}}=(\mueff^2-\mu^2)/\beta$, and equals
\begin{equation}\label{Eq: NP grand potential}
\begin{aligned}
		J_{\text{np}} =&\, -\frac{2\mu}{\beta}\sum_{m=1}^\infty\frac{1}{m}\binom{2m-1}{m-1}\rme^{-2m\mu} +\frac{1}{\beta} \left( \sum_{m=1}^{\infty}\frac{1}{m}\binom{2m-1}{m-1}\rme^{-2m\mu} \right)^2 \\
	=&\, \frac{1}{\beta}\left[-2\mu\rme^{-2\mu} + (1-3\mu)\rme^{-4\mu} + \left(3-\frac{20\mu}{3}\right)\rme^{-6\mu} + \mathcal O(\mu\rme^{-8\mu})\right] \,,
\end{aligned}
\end{equation}
which are two simple towers of $\rme^{-2m\mu}$ that only differ by a polynomial factor of $\mu$. Transforming back to fixed flux with \eqref{Eq: IIB fixed flux} gives a considerably less economical answer,
\begin{equation}\label{Eq: giant graviton form}
	Z(N,\beta) =\rme^{-\frac{\beta N^2}{4}} \mathcal A_{\text{KK}}(\beta) \left(1+\sum_{m\geq1}\frac{(-1)^m(N+2m)}{m}\binom{N+m-1}{m-1}\,q^{m(N+m)}\right)\,,
\end{equation}
in which the grading $q^{m(N+m)}$ is the one expected of $m$ giant gravitons \cite{Imamura:2021ytr,Arai:2020qaj,Gaiotto:2021xce,Beccaria:2024szi}.  Isolating the non-perturbative part the instanton sectors contribute
\begin{equation}\label{Eq: fixed N free energy}
	\log Z_{\text{np}}(N,\beta) =\,-(N+2)\,q^{N+1} - \frac{(N+2)^2}{2}\,q^{2N+2} +\frac{(N+1)(N+4)}{2}\,q^{2N+4} + \mathcal O(q^{3N})\,.
\end{equation}
The leading term was reproduced explicitly by a semi-classical quantisation of a D3-brane giant graviton \cite{Beccaria:2024vfx,Gautason:2024nru}.  As we saw above, in the grand canonical ensemble the same corrections can be reorganised as a renormalisation $\mu\to\mueff$ of the source \eqref{Eq: exact BDF grand index}, collapsing into the two towers of \eqref{Eq: NP grand potential}, of degree one and degree zero in $\mu$. At fixed rank the same content is far less economical, the polynomial degrees in $N$ go up to order $m$ for wrapping number $m$.  A brane computation at fixed flux has to generate that $N$-dependence non-trivially within the brane path integral. One conceivable way for this to happen is through a growing number of unpaired bosonic zero modes whose collective coordinate integrals supply the powers of $N$, while at fixed $c_4^{\mathrm{el}}$ there is nothing of the sort to generate.

That brane instantons can be reabsorbed into a non-perturbative renormalisation of the chemical potential $\mu_{\text{eff}}$ was also seen in M-theory in the context of membrane instantons in the grand canonical partition function of the ABJM theory. In that case the perturbative grand potential is cubic, $J^{\text{pert}}=\frac{\mathcal C}{3}\mu^3+\mathcal B\mu+\mathcal A$, and the membrane bound state contributions are absorbed into an effective chemical potential \cite{Hatsuda:2012dt,Hatsuda:2013gj,Hatsuda:2013oxa},
\begin{equation}\label{Eq: ABJM mueff}
	\mueff=\mu+\frac{1}{C(k)}\sum_{\ell\geq1}a_\ell(k)\,\rme^{-2\ell\mu}\,,
	\qquad
	J(\mu,k)=J^{\text{pert}}(\mueff)+J^{\text{WS}}(\mueff)+J^{\text{M2}}(\mueff)\,,
\end{equation}
after which the worldsheet and membrane instanton towers survive separately. The Schur case is the same mechanism in a sharper form. The perturbative answer evaluated on $\mueff$ provides the exact answer with no more brane instantons left. Grand canonical ensembles in M-theory on AdS$_4\times S^7$ were studied in \cite{Gautason:2025plx,Gautason:2025per} where it was also argued that the semi-classical quantisation of non-degenerate membranes in the bulk were to be done directly in that ensemble and are one-loop exact. It is still to be established if D3-brane giants turn out to be much simpler, at the quantum level, at fixed $c_4^{\mathrm{el}}$, compared to the fixed flux ensemble.
\section{Future directions}
We have focussed on the Schur index of $\mathcal N=4$ SYM because it is under good control in both the canonical and the grand canonical ensemble of the boundary theory, and thus provides a clean case study with which to initiate a broader study of consistent holographic ensembles in type IIB string theory and their semi-classical limits. We found that the ensemble is chosen by a choice of boundary polarizations of the self-dual gauge potential, which is fixed by the coupling of the topological term in the IIB action. Interesting future studies for holographic ensembles in type IIB extend these results to more involved observables and backgrounds.

The most immediate direction is more observables of the same kind. Indices with a known Fermi-gas description include the 1/2-BPS index \cite{Berenstein:2004kk} and the Schur index of the $\mathcal N=2^*$ theory \cite{Hatsuda:2022xdv}. Both would supply additional data on ensemble structures in type IIB, and the latter comes with extra non-trivial bulk fluxes, for which boundary conditions have to be chosen and the topological term of \cite{Kurlyand:2022vzv} has to be complemented with the additional terms discussed in \cite{Adhikari:2026rfb}. On the same footing are the S-fold partition functions of $\mathcal N=4$ \cite{Assel:2018vtq}, which lie on an $\mathcal N=2$ preserving conformal manifold whose type IIB backgrounds were constructed in \cite{Bobev:2023bxs}. These geometries are considerably more involved and have an asymptotic AdS$_4$ boundary. Nevertheless, the boundary partition function is under good control in both the canonical and grand canonical ensemble and as such these backgrounds are an ideal next case study for type IIB holographic ensembles.

Ultimately it would be interesting to test whether holographic ensembles in type IIB string theory are as powerful in resumming $1/N$ corrections as they were for AdS$_4$ backgrounds in M-theory \cite{Gautason:2025plx}. The task is thus to test the IIB ensembles on observables more involved than indices. The expectation value of the maximally supersymmetric Wilson line in the fundamental representation of $\mathcal N=4$ is possibly the prime example. It is known analytically in $N$ and $\lambda$ as a Laguerre polynomial \cite{Drukker:2000rr}, while the corresponding grand canonical answer collapses to a single exponential. It would be interesting to see whether that exponential can be produced directly in the bulk, in which case the entire $1/N$ structure of the Laguerre polynomial would again be an artefact of the ensemble changing integral rather than of the bulk dynamics. This structure is in fact very much analogous to the Wilson line expectation values in 3d $\mathcal N=2$ CS matter theories which are argued to be one-loop exact in the grand canonical ensemble, but have an Airy function structure in the canonical ensemble \cite{Gautason:2025plx,Gautason:2025bft}.\footnote{To prove the one-loop exactness it would be interesting to study the supersymmetry preserved on the worldvolume of the branes extensively, similar to the discussion in \cite{Kurlyand:2026yke}.}

A similar version of the same question concerns local correlation functions rather than partition functions or single-operator expectation values. In the M2-brane theories the protected sphere correlators of the one-dimensional topological sector \cite{Dedushenko:2016jxl} acquire a free Fermi-gas presentation once they are placed in a grand canonical ensemble, and it was observed in \cite{Gaiotto:2020vqj} that their large-$\mu$ expansion truncates, they are polynomial in $\mu$ with no $1/\mu$ tail. That truncation can now be understood as a consequence of the fixed $A_3$- or $\mu$-ensemble in the bulk and it would be interesting to understand what the connection between this holographic ensemble and the twisted M-theory proposal of \cite{Gaiotto:2020vqj} is. Whether analogous simplifications exist in the IIB setup remains to be studied. In this regard it similarly would be interesting to understand how the holographic description of the chiral algebra of $\mathcal N=4$ SYM \cite{Beem:2013sza}, the twisted holographic B-model in \cite{Costello:2018zrm}, is related to the fixed $C_4$- or $\mu$-ensemble in the bulk.

Finally, there are observables for which the boundary side is not yet under control. There is no known Fermi description of the $\mathcal N=2^*$ sphere partition function \cite{Pestun:2007rz}, and to our knowledge no study of its grand canonical counterpart, the IIB geometry is known \cite{Bobev:2018hbq}, which makes it a natural next step. Even less is available for the $\mathcal N=1^*$ deformation, whose sphere partition function is not amenable to supersymmetric localisation but the bulk does provide access to its strongly coupled observables \cite{Bobev:2016nua} and the IIB backgrounds have been constructed explicitly \cite{Bobev:2018eer,Petrini:2018pjk,Bobev:2019wnf}. In both cases identifying the consistent set of boundary conditions, and in particular which sign of the topological term is compatible with the fluxes that are switched on, would be the first step towards a grand canonical description. 

Having a first principled derivation of the consistent boundary polarizations is outside the scope of this work. It would have to come from a direct derivation of the IIB supergravity action from a one-loop quantisation of the degenerate genus zero IIB string \cite{Fradkin:1985fq,Fradkin:1984pq,Tseytlin:1988tv}, including the boundary terms. Attempts regarding the GHY term were made in \cite{Ahmadain:2024uyo,Ahmadain:2024uom} but we are not aware of similar attempts regarding the topological term of \cite{Kurlyand:2022vzv}.

\section*{Acknowledgements}
I am grateful to Fridrik Freyr Gautason and Paul Luis Roehl for useful discussions and to Arkady Tseytlin for comments on the draft. I am supported by the STFC Consolidated Grant ST/X000575/1 and am grateful for the continuous hospitality of the ITF at the KU Leuven.

\appendix
\section{The Schur Casimir prefactor}\label{App: casimir}

This appendix fixes the normalization relating the Schur index to the partition function on $S^1\times S^3$.  We keep the discussion self-contained, since the factor in question controls the leading large-$N$ grand potential.  Let $q=\rme^{-\beta}$ and denote the $U(N)$ Schur index by $\mathcal I_N(q)$ \cite{Bourdier:2015wda}. The supersymmetric partition function is
\begin{equation}\label{Eq: appendix index partition function}
	 Z_N(\beta) = \rme^{-\beta E_{\rm susy}(N)}\mathcal I_N(q) = q^{N^2/4}\mathcal I_N(q)\,,
	\qquad E_{\rm susy}(N)=\frac{N^2}{4}\,.
\end{equation}
This is not a harmless overall normalization.  More generally, if $Z_N=q^{cN^2}\mathcal I_N$ and $\mathcal I_N$ tends to an $N$-independent limit up to corrections exponentially small in $N$, then the sum is controlled by a saddle at continuous $N$,
\begin{equation}\label{Eq: appendix grand saddle}
	\widehat\Xi(\mu,\beta)
	=\sum_{N\geq0}\rme^{\mu N} Z_N(\beta)\,,
	\qquad
	N_\ast=\frac{\mu}{2c\beta}\,,
	\qquad
	\log\widehat\Xi=\frac{\mu^2}{4c\beta}+O(\mu^0)\,.
\end{equation}
Thus $c=1/4$ gives the quadratic term $\mu^2/\beta$. We now review how this value is uniquely fixed from the $S^3$ spectrum and localization, and then match it to the holographic renormalization scheme.

\subsection{Regulated mode sum}

We want to compute the contribution to the supersymmetric Casimir energy of an adjoint $\mathcal N=4$ multiplet on the round $S^3$, with the twisted Hamiltonian
\begin{equation}\label{Eq: appendix twisted Hamiltonian}
	\mathcal H_a=E+a\,j_1\,.
\end{equation}
The Schur limit is at $a=1$, which is where the twist \eqref{Eq: Schur twist} sits on the supersymmetric family \eqref{Eq: susy twist family} and at other $a$ the grading is a regulator in a non-supersymmetric background. To perform the mode sums we define the $SU(2)$ character $\chi_j(y)=\sum_{m=-j}^{j}y^m$.  The $S^3$ harmonics fall into representations $(j_L,j_R)$ of $SO(4)=SU(2)_L\times SU(2)_R$ and $j_1=J^3_L+J^3_R$, so the same fugacity $y$ is assigned to both factors; the two vector towers $(\tfrac{k+2}2,\tfrac k2)$ and $(\tfrac k2,\tfrac{k+2}2)$ then coincide, which is the factor of two below.  The single-particle partition functions are
\begin{align}\label{Eq: appendix mode towers}
	z_s(x,y)&=\sum_{k=0}^{\infty}x^{k+1}\chi_{k/2}(y)^2\,,\nonumber\\
	z_f(x,y)&=\sum_{k=0}^{\infty}x^{k+3/2}
	\chi_{(k+1)/2}(y)\chi_{k/2}(y)\,,\\
	z_v(x,y)&=2\sum_{k=0}^{\infty}x^{k+2}
	\chi_{(k+2)/2}(y)\chi_{k/2}(y)\,,\nonumber
\end{align}
and adding them up with correct multiplicities and fermion signs we have for the $\mathcal N=4$ multiplet
\begin{equation}\label{Eq: appendix net letter}
	z_{\mathcal N=4}(x,y)=6z_s(x,y)+z_v(x,y)-8z_f(x,y)\,.
\end{equation}
The standard, twist-blind prescription regulates with the untwisted energy, that is at $y=1$.  The $j_1$-weighted sum then cancels at every level of \eqref{Eq: appendix mode towers} by the $m\leftrightarrow-m$ symmetry of the characters, so the twist drops out altogether and the same answer is returned for every $a$.  What is left is $\tfrac12\sum(-1)^F E$ over the towers, and a $\zeta$-regulator gives
\begin{align}\label{Eq: appendix blind sum}
	E_s&=\frac{6}{2}\zeta(-3)=\frac{1}{40}\,,\nonumber\\
	E_v&=\frac12\bigl[2\zeta(-3)-2\zeta(-1)\bigr]
	=\frac{11}{120}\,,\\
	E_f&=-\frac12\bigl[8\zeta(-3,\tfrac32)
	-2\zeta(-1,\tfrac32)\bigr]=\frac{17}{240}\,,\nonumber
\end{align}
and hence $E_s+E_v+E_f=3/16$, the ordinary Casimir energy, independently of $a$.

The supersymmetric prescription instead sums with the charges appearing in the Hamiltonian \cite{Assel:2015nca}. After resumming the towers in \eqref{Eq: appendix mode towers}, its finite part is
\begin{equation}\label{Eq: appendix Ec}
	E_c(a)
	=-\frac12\underset{\delta\to0}{\operatorname{FP}}
	\frac{\partial}{\partial\delta}
	z_{\mathcal N=4}(\rme^{-\delta},\rme^{-a\delta})
	=\frac{3+a^2}{16}\,.
\end{equation}
Here $\operatorname{FP}$ means minimal subtraction of the divergent powers of $\delta$.  At $a=0$ it reduces to $3/16$, reproducing \eqref{Eq: appendix blind sum}.  At the supersymmetric endpoint $a=1$ the Schur limit of the $\mathcal N=4$ single letter equals
\begin{equation}\label{Eq: appendix Schur letter}
	z_{\mathcal N=4}^{\rm Schur}(q)
	=\lim_{u\to1}z_{\mathcal N=4}(q,qu)
	=\frac{2q}{1+q}\,,
	\quad \text{and} \quad
	E_c(1)=-\frac12\left.
	\frac{\partial}{\partial\delta}
	z_{\mathcal N=4}^{\rm Schur}(\rme^{-\delta})
	\right|_{\delta=0}=\frac14\,.
\end{equation}
Multiplication by $\dim U(N)=N^2$ therefore yields the exponent in \eqref{Eq: appendix index partition function}.  The difference from the untwisted result is the finite term $E_c(a)-E_c(0)=a^2/16$.

\subsection{Localization on the supersymmetric branch}

The same normalization follows from the one-loop-exact prefactor of the localized $\mathcal N=4$ partition function \cite{Cabo-Bizet:2018ehj,Bobev:2015kza,Assel:2015nca}.  In chemical-potential variables it reads
\begin{equation}\label{Eq: appendix CCMM}
	Z_N=\rme^{-\mathcal F}\mathcal I_N\,,
	\qquad
	\mathcal F=\frac{N^2}{2}
	\frac{\Delta_1\Delta_2\Delta_3}{\omega_1\omega_2}\,,
	\qquad
	\sum_{I=1}^{3}\Delta_I-\sum_{i=1}^{2}\omega_i
	=2\pi\rmi n\,,\quad n\in\Z\,.
\end{equation}
The chemical potentials are logarithms of fugacities, so their shifts by integer multiples of $2\pi\rmi$ and the integer $n$ must be specified together.  The supersymmetric Casimir energy belongs to the $n=0$ branch \cite{Cabo-Bizet:2018ehj}.  On this branch the Schur twist \eqref{Eq: Schur twist} gives, through the dictionary $\omega_a=\beta(1+\varepsilon_a)$ and $\Delta_I=\beta(1-\sigma_I)$ of section~\ref{sec: schur background},
\begin{equation}\label{Eq: appendix physical Schur representative}
	(\Delta_1,\Delta_2,\Delta_3;\omega_1,\omega_2)
	=(\beta,\beta,\beta;2\beta,\beta)\,,
	\qquad n=0\,.
\end{equation}
The corresponding single-letter function and the prefactor are, respectively,
\begin{equation}\label{Eq: appendix localized letter}
	f(\Delta_I,\omega_i)
	=1-\frac{\prod_{I=1}^{3}(1-\rme^{-\Delta_I})}
	{\prod_{i=1}^{2}(1-\rme^{-\omega_i})}
	=\frac{2q}{1+q}\,,\quad
	\mathcal F
	=\frac{N^2}{2}\frac{\beta^3}{(2\beta)\beta}
	=\frac{\beta N^2}{4}\,.
\end{equation}
This reproduces \eqref{Eq: appendix index partition function} without a mode sum.  The representative is not a choice of convenience.  With the internal twists off the three $\Delta_I$ are equal, and on the $n=0$ branch either the Schur letter $f=2q/(1+q)$ or the value $\mathcal F=\beta N^2/4$ then forces $\{\omega_1,\omega_2\}=\{2\beta,\beta\}$, the two solutions differing only by which plane carries the twist.  This is the complex-structure ratio $b_1/b_2=2$ of \eqref{Eq: Schur moduli}.  

\subsection{Holographic renormalization scheme}

The field-theory shift has a direct holographic counterpart.  Standard covariant counterterms are blind to the twist in the same way the $y=1$ mode sum is: the minimal renormalised volume \eqref{Eq: minimal volume} gives the ordinary Casimir contribution $I_{\rm standard}=3\beta N^2/16$ on the Schur background just as on the untwisted one.  Supersymmetric holographic renormalization instead adds the finite boundary terms required by the supersymmetric Ward identities \cite{BenettiGenolini:2016tsn}.  For the twisted background with moduli $(b_1,b_2)$, the resulting on-shell action is
\begin{equation}\label{Eq: appendix supersymmetric action}
	I_{\rm susy}
	=\frac{2\beta}{27}
	\frac{(b_1+b_2)^3}{b_1b_2}
	\frac{\pi^2L^3}{\kappa_5^2}\,,
	\qquad
	\frac{\pi^2L^3}{\kappa_5^2}=\frac{L^8}{64\pi^2\ell_s^8}\,.
\end{equation}
The second relation in \eqref{Eq: appendix supersymmetric action} is the standard $\text{AdS}_5\times S^5$ consistent-truncation normalization where the five-dimensional coupling is computed by evaluating the integral over the internal $S^5$.  At the Schur point $(b_1,b_2)=(2,1)$ this becomes
\begin{equation}\label{Eq: appendix holographic Schur action}
	I_{\rm susy}=\frac{\beta N^2}{4}\,,
	\qquad
	I_{\rm susy}-I_{\rm standard}=\frac{\beta N^2}{16}\,,
\end{equation}
where we used \eqref{Eq: flux and chemical potential}.  The two schemes reproduce the two mode sums separately, $I_{\rm susy}=\beta N^2E_c(1)$ and $I_{\rm standard}=\beta N^2E_c(0)$. Consequently the factor $q^{N^2/4}$ is fixed consistently by the Schur spectrum, localization, and the supersymmetric bulk counterterm scheme, it is the physical Casimir prefactor that produces the quadratic grand potential in \eqref{Eq: appendix grand saddle}.

\bibliography{IIBensembles.bib}

\providecommand{\href}[2]{#2}\begingroup\raggedright\begin{thebibliography}{10}

\bibitem{Maldacena:1997re}
J.~M. Maldacena, \emph{{The Large $N$ limit of superconformal field theories and supergravity}}, \href{http://dx.doi.org/10.4310/ATMP.1998.v2.n2.a1}{\emph{Adv. Theor. Math. Phys.} {\bf 2} (1998) 231--252}, [\href{https://arxiv.org/abs/hep-th/9711200}{{\tt hep-th/9711200}}].

\bibitem{Gubser:1998bc}
S.~S. Gubser, I.~R. Klebanov and A.~M. Polyakov, \emph{{Gauge theory correlators from noncritical string theory}}, \href{http://dx.doi.org/10.1016/S0370-2693(98)00377-3}{\emph{Phys. Lett. B} {\bf 428} (1998) 105--114}, [\href{https://arxiv.org/abs/hep-th/9802109}{{\tt hep-th/9802109}}].

\bibitem{Witten:1998qj}
E.~Witten, \emph{{Anti de Sitter space and holography}}, \href{http://dx.doi.org/10.4310/ATMP.1998.v2.n2.a2}{\emph{Adv. Theor. Math. Phys.} {\bf 2} (1998) 253--291}, [\href{https://arxiv.org/abs/hep-th/9802150}{{\tt hep-th/9802150}}].

\bibitem{Gautason:2025plx}
F.~F. Gautason and J.~van Muiden, \emph{{Ensembles in M-theory and holography}}, \href{http://dx.doi.org/10.1007/JHEP11(2025)078}{\emph{JHEP} {\bf 11} (2025) 078}, [\href{https://arxiv.org/abs/2505.21633}{{\tt 2505.21633}}].

\bibitem{Gautason:2025per}
F.~F. Gautason and J.~van Muiden, \emph{{Localization of the M2-Brane}}, \href{http://dx.doi.org/10.1103/67bh-xd42}{\emph{Phys. Rev. Lett.} {\bf 135} (2025) 101601}, [\href{https://arxiv.org/abs/2503.16597}{{\tt 2503.16597}}].

\bibitem{Beccaria:2023ujc}
M.~Beccaria, S.~Giombi and A.~A. Tseytlin, \emph{{Instanton contributions to the ABJM free energy from quantum M2 branes}}, \href{http://dx.doi.org/10.1007/JHEP10(2023)029}{\emph{JHEP} {\bf 10} (2023) 029}, [\href{https://arxiv.org/abs/2307.14112}{{\tt 2307.14112}}].

\bibitem{Gautason:2023igo}
F.~F. Gautason, V.~G.~M. Puletti and J.~van Muiden, \emph{{Quantized strings and instantons in holography}}, \href{http://dx.doi.org/10.1007/JHEP08(2023)218}{\emph{JHEP} {\bf 08} (2023) 218}, [\href{https://arxiv.org/abs/2304.12340}{{\tt 2304.12340}}].

\bibitem{Bobev:2026gir}
N.~Bobev, F.~F. Gautason and J.~van Muiden, \emph{{Holographic Tests of the $\mu$ Ensemble}},  \href{https://arxiv.org/abs/2607.06493}{{\tt 2607.06493}}.

\bibitem{Aharony:2008ug}
O.~Aharony, O.~Bergman, D.~L. Jafferis and J.~Maldacena, \emph{{N=6 superconformal Chern-Simons-matter theories, M2-branes and their gravity duals}}, \href{http://dx.doi.org/10.1088/1126-6708/2008/10/091}{\emph{JHEP} {\bf 10} (2008) 091}, [\href{https://arxiv.org/abs/0806.1218}{{\tt 0806.1218}}].

\bibitem{Marino:2011eh}
M.~Marino and P.~Putrov, \emph{{ABJM theory as a Fermi gas}}, \href{http://dx.doi.org/10.1088/1742-5468/2012/03/P03001}{\emph{J. Stat. Mech.} {\bf 1203} (2012) P03001}, [\href{https://arxiv.org/abs/1110.4066}{{\tt 1110.4066}}].

\bibitem{Hatsuda:2012hm}
Y.~Hatsuda, S.~Moriyama and K.~Okuyama, \emph{{Exact Results on the ABJM Fermi Gas}}, \href{http://dx.doi.org/10.1007/JHEP10(2012)020}{\emph{JHEP} {\bf 10} (2012) 020}, [\href{https://arxiv.org/abs/1207.4283}{{\tt 1207.4283}}].

\bibitem{Fuji:2011km}
H.~Fuji, S.~Hirano and S.~Moriyama, \emph{{Summing Up All Genus Free Energy of ABJM Matrix Model}}, \href{http://dx.doi.org/10.1007/JHEP08(2011)001}{\emph{JHEP} {\bf 08} (2011) 001}, [\href{https://arxiv.org/abs/1106.4631}{{\tt 1106.4631}}].

\bibitem{BenettiGenolini:2026cyc}
P.~Benetti~Genolini, F.~Gaar, J.~P. Gauntlett, J.~Park and J.~Sparks, \emph{{Airy functions from quantum M-theory}},  \href{https://arxiv.org/abs/2607.07255}{{\tt 2607.07255}}.

\bibitem{Bhattacharyya:2012ye}
S.~Bhattacharyya, A.~Grassi, M.~Marino and A.~Sen, \emph{{A One-Loop Test of Quantum Supergravity}}, \href{http://dx.doi.org/10.1088/0264-9381/31/1/015012}{\emph{Class. Quant. Grav.} {\bf 31} (2014) 015012}, [\href{https://arxiv.org/abs/1210.6057}{{\tt 1210.6057}}].

\bibitem{Gadde:2011ik}
A.~Gadde, L.~Rastelli, S.~S. Razamat and W.~Yan, \emph{{The 4d Superconformal Index from q-deformed 2d Yang-Mills}}, \href{http://dx.doi.org/10.1103/PhysRevLett.106.241602}{\emph{Phys. Rev. Lett.} {\bf 106} (2011) 241602}, [\href{https://arxiv.org/abs/1104.3850}{{\tt 1104.3850}}].

\bibitem{Gadde:2011uv}
A.~Gadde, L.~Rastelli, S.~S. Razamat and W.~Yan, \emph{{Gauge Theories and Macdonald Polynomials}}, \href{http://dx.doi.org/10.1007/s00220-012-1607-8}{\emph{Commun. Math. Phys.} {\bf 319} (2013) 147--193}, [\href{https://arxiv.org/abs/1110.3740}{{\tt 1110.3740}}].

\bibitem{Bourdier:2015wda}
J.~Bourdier, N.~Drukker and J.~Felix, \emph{{The exact Schur index of $\mathcal{N}=4$ SYM}}, \href{http://dx.doi.org/10.1007/JHEP11(2015)210}{\emph{JHEP} {\bf 11} (2015) 210}, [\href{https://arxiv.org/abs/1507.08659}{{\tt 1507.08659}}].

\bibitem{Kurlyand:2022vzv}
S.~A. Kurlyand and A.~A. Tseytlin, \emph{{Type IIB supergravity action on $M_5 \times X_5$ solutions}}, \href{http://dx.doi.org/10.1103/PhysRevD.106.086017}{\emph{Phys. Rev. D} {\bf 106} (2022) 086017}, [\href{https://arxiv.org/abs/2206.14522}{{\tt 2206.14522}}].

\bibitem{Fradkin:1985fq}
E.~S. Fradkin and A.~A. Tseytlin, \emph{{Effective Action Approach to Superstring Theory}}, \href{http://dx.doi.org/10.1016/0370-2693(85)91468-6}{\emph{Phys. Lett. B} {\bf 160} (1985) 69--76}.

\bibitem{Fradkin:1984pq}
E.~S. Fradkin and A.~A. Tseytlin, \emph{{Effective Field Theory from Quantized Strings}}, \href{http://dx.doi.org/10.1016/0370-2693(85)91190-6}{\emph{Phys. Lett. B} {\bf 158} (1985) 316--322}.

\bibitem{Tseytlin:1988tv}
A.~A. Tseytlin, \emph{{Mobius Infinity Subtraction and Effective Action in $\sigma$ Model Approach to Closed String Theory}}, \href{http://dx.doi.org/10.1016/0370-2693(88)90421-2}{\emph{Phys. Lett. B} {\bf 208} (1988) 221--227}.

\bibitem{Gunaydin:1984fk}
M.~Gunaydin and N.~Marcus, \emph{{The Spectrum of the s**5 Compactification of the Chiral N=2, D=10 Supergravity and the Unitary Supermultiplets of U(2, 2/4)}}, \href{http://dx.doi.org/10.1088/0264-9381/2/2/001}{\emph{Class. Quant. Grav.} {\bf 2} (1985) L11}.

\bibitem{Kim:1985ez}
H.~Kim, L.~Romans and P.~van Nieuwenhuizen, \emph{{The Mass Spectrum of Chiral N=2 D=10 Supergravity on S**5}}, \href{http://dx.doi.org/10.1103/PhysRevD.32.389}{\emph{Phys. Rev. D} {\bf 32} (1985) 389}.

\bibitem{Aharony:1998qu}
O.~Aharony and E.~Witten, \emph{{Anti-de Sitter space and the center of the gauge group}}, \href{http://dx.doi.org/10.1088/1126-6708/1998/11/018}{\emph{JHEP} {\bf 11} (1998) 018}, [\href{https://arxiv.org/abs/hep-th/9807205}{{\tt hep-th/9807205}}].

\bibitem{Gaiotto:2021xce}
D.~Gaiotto and J.~H. Lee, \emph{{The giant graviton expansion}}, \href{http://dx.doi.org/10.1007/JHEP08(2024)025}{\emph{JHEP} {\bf 08} (2024) 025}, [\href{https://arxiv.org/abs/2109.02545}{{\tt 2109.02545}}].

\bibitem{Imamura:2021ytr}
Y.~Imamura, \emph{{Finite-N superconformal index via the AdS/CFT correspondence}}, \href{http://dx.doi.org/10.1093/ptep/ptab141}{\emph{PTEP} {\bf 2021} (2021) 123B05}, [\href{https://arxiv.org/abs/2108.12090}{{\tt 2108.12090}}].

\bibitem{Arai:2020qaj}
R.~Arai, S.~Fujiwara, Y.~Imamura and T.~Mori, \emph{{Schur index of the ${\cal N}=4$ $U(N)$ supersymmetric Yang-Mills theory via the AdS/CFT correspondence}}, \href{http://dx.doi.org/10.1103/PhysRevD.101.086017}{\emph{Phys. Rev. D} {\bf 101} (2020) 086017}, [\href{https://arxiv.org/abs/2001.11667}{{\tt 2001.11667}}].

\bibitem{Beccaria:2024vfx}
M.~Beccaria and A.~Cabo-Bizet, \emph{{Large $N$ Schur index of $\mathcal N=4$ SYM from semiclassical D3 brane}},  \href{https://arxiv.org/abs/2402.12172}{{\tt 2402.12172}}.

\bibitem{Gautason:2024nru}
F.~F. Gautason and J.~van Muiden, \emph{{One-loop quantization of Euclidean D3-branes in holographic backgrounds}}, \href{http://dx.doi.org/10.1007/JHEP06(2024)073}{\emph{JHEP} {\bf 06} (2024) 073}, [\href{https://arxiv.org/abs/2402.16779}{{\tt 2402.16779}}].

\bibitem{Pasti:1996vs}
P.~Pasti, D.~P. Sorokin and M.~Tonin, \emph{{On Lorentz invariant actions for chiral p forms}}, \href{http://dx.doi.org/10.1103/PhysRevD.55.6292}{\emph{Phys. Rev. D} {\bf 55} (1997) 6292--6298}, [\href{https://arxiv.org/abs/hep-th/9611100}{{\tt hep-th/9611100}}].

\bibitem{DallAgata:1997gnw}
G.~Dall'Agata, K.~Lechner and D.~P. Sorokin, \emph{{Covariant actions for the bosonic sector of d = 10 IIB supergravity}}, \href{http://dx.doi.org/10.1088/0264-9381/14/12/003}{\emph{Class. Quant. Grav.} {\bf 14} (1997) L195--L198}, [\href{https://arxiv.org/abs/hep-th/9707044}{{\tt hep-th/9707044}}].

\bibitem{DallAgata:1998ahf}
G.~Dall'Agata, K.~Lechner and M.~Tonin, \emph{{D = 10, N = IIB supergravity: Lorentz invariant actions and duality}}, \href{http://dx.doi.org/10.1088/1126-6708/1998/07/017}{\emph{JHEP} {\bf 07} (1998) 017}, [\href{https://arxiv.org/abs/hep-th/9806140}{{\tt hep-th/9806140}}].

\bibitem{Adhikari:2026rfb}
S.~Adhikari, J.~Hong, C.~Joung and G.~Lee, \emph{{Type IIB supergravity action and holography}}, \href{http://dx.doi.org/10.1007/JHEP06(2026)271}{\emph{JHEP} {\bf 06} (2026) 271}, [\href{https://arxiv.org/abs/2603.18248}{{\tt 2603.18248}}].

\bibitem{vanMuiden:2026nsp}
J.~van Muiden, \emph{{Quantum M2-branes and Holography}},  \href{https://arxiv.org/abs/2603.14544}{{\tt 2603.14544}}.

\bibitem{Assel:2014paa}
B.~Assel, D.~Cassani and D.~Martelli, \emph{{Localization on Hopf surfaces}}, \href{http://dx.doi.org/10.1007/JHEP08(2014)123}{\emph{JHEP} {\bf 08} (2014) 123}, [\href{https://arxiv.org/abs/1405.5144}{{\tt 1405.5144}}].

\bibitem{Henningson:1998gx}
M.~Henningson and K.~Skenderis, \emph{{The Holographic Weyl anomaly}}, \href{http://dx.doi.org/10.1088/1126-6708/1998/07/023}{\emph{JHEP} {\bf 07} (1998) 023}, [\href{https://arxiv.org/abs/hep-th/9806087}{{\tt hep-th/9806087}}].

\bibitem{Emparan:1999pm}
R.~Emparan, C.~V. Johnson and R.~C. Myers, \emph{{Surface terms as counterterms in the AdS / CFT correspondence}}, \href{http://dx.doi.org/10.1103/PhysRevD.60.104001}{\emph{Phys. Rev. D} {\bf 60} (1999) 104001}, [\href{https://arxiv.org/abs/hep-th/9903238}{{\tt hep-th/9903238}}].

\bibitem{Skenderis:2002wp}
K.~Skenderis, \emph{{Lecture notes on holographic renormalization}}, \href{http://dx.doi.org/10.1088/0264-9381/19/22/306}{\emph{Class. Quant. Grav.} {\bf 19} (2002) 5849--5876}, [\href{https://arxiv.org/abs/hep-th/0209067}{{\tt hep-th/0209067}}].

\bibitem{Assel:2015nca}
B.~Assel, D.~Cassani, L.~Di~Pietro, Z.~Komargodski, J.~Lorenzen and D.~Martelli, \emph{{The Casimir Energy in Curved Space and its Supersymmetric Counterpart}}, \href{http://dx.doi.org/10.1007/JHEP07(2015)043}{\emph{JHEP} {\bf 07} (2015) 043}, [\href{https://arxiv.org/abs/1503.05537}{{\tt 1503.05537}}].

\bibitem{BenettiGenolini:2016qwm}
P.~Benetti~Genolini, D.~Cassani, D.~Martelli and J.~Sparks, \emph{{The holographic supersymmetric Casimir energy}}, \href{http://dx.doi.org/10.1103/PhysRevD.95.021902}{\emph{Phys. Rev. D} {\bf 95} (2017) 021902}, [\href{https://arxiv.org/abs/1606.02724}{{\tt 1606.02724}}].

\bibitem{BenettiGenolini:2016tsn}
P.~Benetti~Genolini, D.~Cassani, D.~Martelli and J.~Sparks, \emph{{Holographic renormalization and supersymmetry}}, \href{http://dx.doi.org/10.1007/JHEP02(2017)132}{\emph{JHEP} {\bf 02} (2017) 132}, [\href{https://arxiv.org/abs/1612.06761}{{\tt 1612.06761}}].

\bibitem{Kinney:2005ej}
J.~Kinney, J.~M. Maldacena, S.~Minwalla and S.~Raju, \emph{{An Index for 4 dimensional super conformal theories}}, \href{http://dx.doi.org/10.1007/s00220-007-0258-7}{\emph{Commun. Math. Phys.} {\bf 275} (2007) 209--254}, [\href{https://arxiv.org/abs/hep-th/0510251}{{\tt hep-th/0510251}}].

\bibitem{Romelsberger:2005eg}
C.~Romelsberger, \emph{{Counting chiral primaries in N = 1, d=4 superconformal field theories}}, \href{http://dx.doi.org/10.1016/j.nuclphysb.2006.03.037}{\emph{Nucl. Phys. B} {\bf 747} (2006) 329--353}, [\href{https://arxiv.org/abs/hep-th/0510060}{{\tt hep-th/0510060}}].

\bibitem{Beccaria:2024szi}
M.~Beccaria and A.~Cabo-Bizet, \emph{{Giant graviton expansion of Schur index and quasimodular forms}}, \href{http://dx.doi.org/10.1007/JHEP05(2024)282}{\emph{JHEP} {\bf 05} (2024) 282}, [\href{https://arxiv.org/abs/2403.06509}{{\tt 2403.06509}}].

\bibitem{Hatsuda:2012dt}
Y.~Hatsuda, S.~Moriyama and K.~Okuyama, \emph{{Instanton Effects in ABJM Theory from Fermi Gas Approach}}, \href{http://dx.doi.org/10.1007/JHEP01(2013)158}{\emph{JHEP} {\bf 01} (2013) 158}, [\href{https://arxiv.org/abs/1211.1251}{{\tt 1211.1251}}].

\bibitem{Hatsuda:2013gj}
Y.~Hatsuda, S.~Moriyama and K.~Okuyama, \emph{{Instanton Bound States in ABJM Theory}}, \href{http://dx.doi.org/10.1007/JHEP05(2013)054}{\emph{JHEP} {\bf 05} (2013) 054}, [\href{https://arxiv.org/abs/1301.5184}{{\tt 1301.5184}}].

\bibitem{Hatsuda:2013oxa}
Y.~Hatsuda, M.~Marino, S.~Moriyama and K.~Okuyama, \emph{{Non-perturbative effects and the refined topological string}}, \href{http://dx.doi.org/10.1007/JHEP09(2014)168}{\emph{JHEP} {\bf 09} (2014) 168}, [\href{https://arxiv.org/abs/1306.1734}{{\tt 1306.1734}}].

\bibitem{Berenstein:2004kk}
D.~Berenstein, \emph{{A Toy model for the AdS / CFT correspondence}}, \href{http://dx.doi.org/10.1088/1126-6708/2004/07/018}{\emph{JHEP} {\bf 07} (2004) 018}, [\href{https://arxiv.org/abs/hep-th/0403110}{{\tt hep-th/0403110}}].

\bibitem{Hatsuda:2022xdv}
Y.~Hatsuda and T.~Okazaki, \emph{{$ \mathcal{N} $ = 2$^{*}$ Schur indices}}, \href{http://dx.doi.org/10.1007/JHEP01(2023)029}{\emph{JHEP} {\bf 01} (2023) 029}, [\href{https://arxiv.org/abs/2208.01426}{{\tt 2208.01426}}].

\bibitem{Assel:2018vtq}
B.~Assel and A.~Tomasiello, \emph{{Holographic duals of 3d S-fold CFTs}}, \href{http://dx.doi.org/10.1007/JHEP06(2018)019}{\emph{JHEP} {\bf 06} (2018) 019}, [\href{https://arxiv.org/abs/1804.06419}{{\tt 1804.06419}}].

\bibitem{Bobev:2023bxs}
N.~Bobev, F.~F. Gautason and J.~van Muiden, \emph{{The conformal manifold of S-folds in string theory}}, \href{http://dx.doi.org/10.1007/JHEP03(2024)167}{\emph{JHEP} {\bf 03} (2024) 167}, [\href{https://arxiv.org/abs/2312.13370}{{\tt 2312.13370}}].

\bibitem{Drukker:2000rr}
N.~Drukker and D.~J. Gross, \emph{{An Exact prediction of N=4 SUSYM theory for string theory}}, \href{http://dx.doi.org/10.1063/1.1372177}{\emph{J. Math. Phys.} {\bf 42} (2001) 2896--2914}, [\href{https://arxiv.org/abs/hep-th/0010274}{{\tt hep-th/0010274}}].

\bibitem{Gautason:2025bft}
F.~F. Gautason and A.~Nix, \emph{{Universal holographic Wilson loops in 3d SCFTs}}, \href{http://dx.doi.org/10.1007/JHEP06(2026)061}{\emph{JHEP} {\bf 06} (2026) 061}, [\href{https://arxiv.org/abs/2511.04596}{{\tt 2511.04596}}].

\bibitem{Kurlyand:2026yke}
S.~A. Kurlyand, \emph{{Membrane instantons and non-perturbative effects in $\mathrm{AdS}_{4}/\mathrm{CFT}_{3}$}},  \href{https://arxiv.org/abs/2606.19467}{{\tt 2606.19467}}.

\bibitem{Dedushenko:2016jxl}
M.~Dedushenko, S.~S. Pufu and R.~Yacoby, \emph{{A one-dimensional theory for Higgs branch operators}}, \href{http://dx.doi.org/10.1007/JHEP03(2018)138}{\emph{JHEP} {\bf 03} (2018) 138}, [\href{https://arxiv.org/abs/1610.00740}{{\tt 1610.00740}}].

\bibitem{Gaiotto:2020vqj}
D.~Gaiotto and J.~Abajian, \emph{{Twisted M2 brane holography and sphere correlation functions}}, \href{http://dx.doi.org/10.1007/JHEP03(2025)195}{\emph{JHEP} {\bf 03} (2025) 195}, [\href{https://arxiv.org/abs/2004.13810}{{\tt 2004.13810}}].

\bibitem{Beem:2013sza}
C.~Beem, M.~Lemos, P.~Liendo, W.~Peelaers, L.~Rastelli and B.~C. van Rees, \emph{{Infinite Chiral Symmetry in Four Dimensions}}, \href{http://dx.doi.org/10.1007/s00220-014-2272-x}{\emph{Commun. Math. Phys.} {\bf 336} (2015) 1359--1433}, [\href{https://arxiv.org/abs/1312.5344}{{\tt 1312.5344}}].

\bibitem{Costello:2018zrm}
K.~Costello and D.~Gaiotto, \emph{{Twisted holography}}, \href{http://dx.doi.org/10.1007/JHEP01(2025)087}{\emph{JHEP} {\bf 01} (2025) 087}, [\href{https://arxiv.org/abs/1812.09257}{{\tt 1812.09257}}].

\bibitem{Pestun:2007rz}
V.~Pestun, \emph{{Localization of gauge theory on a four-sphere and supersymmetric Wilson loops}}, \href{http://dx.doi.org/10.1007/s00220-012-1485-0}{\emph{Commun. Math. Phys.} {\bf 313} (2012) 71--129}, [\href{https://arxiv.org/abs/0712.2824}{{\tt 0712.2824}}].

\bibitem{Bobev:2018hbq}
N.~Bobev, F.~F. Gautason and J.~Van~Muiden, \emph{{Precision Holography for $\mathcal{N}=2^{*}$ on $S^4$ from type IIB Supergravity}}, \href{http://dx.doi.org/10.1007/JHEP04(2018)148}{\emph{JHEP} {\bf 04} (2018) 148}, [\href{https://arxiv.org/abs/1802.09539}{{\tt 1802.09539}}].

\bibitem{Bobev:2016nua}
N.~Bobev, H.~Elvang, U.~Kol, T.~Olson and S.~S. Pufu, \emph{{Holography for $ \mathcal{N} = 1^*$ on S$^{4}$}}, \href{http://dx.doi.org/10.1007/JHEP10(2016)095}{\emph{JHEP} {\bf 10} (2016) 095}, [\href{https://arxiv.org/abs/1605.00656}{{\tt 1605.00656}}].

\bibitem{Bobev:2018eer}
N.~Bobev, F.~F. Gautason, B.~E. Niehoff and J.~van Muiden, \emph{{Uplifting GPPZ: a ten-dimensional dual of $ \mathcal{N}={1}^{\ast } $}}, \href{http://dx.doi.org/10.1007/JHEP10(2018)058}{\emph{JHEP} {\bf 10} (2018) 058}, [\href{https://arxiv.org/abs/1805.03623}{{\tt 1805.03623}}].

\bibitem{Petrini:2018pjk}
M.~Petrini, H.~Samtleben, S.~Schmidt and K.~Skenderis, \emph{{The 10d Uplift of the GPPZ Solution}}, \href{http://dx.doi.org/10.1007/JHEP07(2018)026}{\emph{JHEP} {\bf 07} (2018) 026}, [\href{https://arxiv.org/abs/1805.01919}{{\tt 1805.01919}}].

\bibitem{Bobev:2019wnf}
N.~Bobev, F.~F. Gautason, B.~E. Niehoff and J.~van Muiden, \emph{{A holographic kaleidoscope for $ \mathcal{N} $ = 1*}}, \href{http://dx.doi.org/10.1007/JHEP10(2019)185}{\emph{JHEP} {\bf 10} (2019) 185}, [\href{https://arxiv.org/abs/1906.09270}{{\tt 1906.09270}}].

\bibitem{Ahmadain:2024uyo}
A.~Ahmadain, S.~Akhtar and R.~Khan, \emph{{The GHY boundary term from the string worldsheet to linear order}},  \href{https://arxiv.org/abs/2411.06400}{{\tt 2411.06400}}.

\bibitem{Ahmadain:2024uom}
A.~Ahmadain, V.~Shyam and Z.~Yan, \emph{{A Comment on Deriving the Gibbons-Hawking-York Term From the String Worldsheet}},  \href{https://arxiv.org/abs/2407.18866}{{\tt 2407.18866}}.

\bibitem{Cabo-Bizet:2018ehj}
A.~Cabo-Bizet, D.~Cassani, D.~Martelli and S.~Murthy, \emph{{Microscopic origin of the Bekenstein-Hawking entropy of supersymmetric AdS$_{5}$ black holes}}, \href{http://dx.doi.org/10.1007/JHEP10(2019)062}{\emph{JHEP} {\bf 10} (2019) 062}, [\href{https://arxiv.org/abs/1810.11442}{{\tt 1810.11442}}].

\bibitem{Bobev:2015kza}
N.~Bobev, M.~Bullimore and H.-C. Kim, \emph{{Supersymmetric Casimir Energy and the Anomaly Polynomial}}, \href{http://dx.doi.org/10.1007/JHEP09(2015)142}{\emph{JHEP} {\bf 09} (2015) 142}, [\href{https://arxiv.org/abs/1507.08553}{{\tt 1507.08553}}].

\end{thebibliography}\endgroup
\bibliographystyle{JHEP}

\end{document}